\RequirePackage{fix-cm}
\documentclass[smallextended]{svjour3_nobox} 

\usepackage{graphicx}  

\smartqed  

\usepackage{amsmath}
\usepackage{paralist}

\usepackage{amsfonts} 
\usepackage{amssymb} 
\usepackage{dcolumn}
\usepackage{bm}
\usepackage{mathrsfs}
\usepackage{newtxtext} 
\usepackage{newtxmath} 
\usepackage{relsize}
\usepackage{xspace}
\usepackage[per-mode=symbol,range-phrase = \text{--}]{siunitx}
\usepackage[cal=boondoxo]{mathalfa}
\usepackage[defaultlines=2,all]{nowidow}
\usepackage{calc}
\usepackage{physics}
\usepackage{dynkin-diagrams}
\usepackage{layout}

\newcommand{\vva}[1]{\va*{#1}}

\newcommand{\gr}[2]{\ensuremath{\mathbf{#1}_{#2}}}
\newcommand{\rnk}{\ensuremath{\ell}\xspace}
\DeclareMathOperator{\proj}{Pr}
\newcommand{\nrm}[1]{|#1|}
\usepackage[pagewise]{lineno}

\begin{document}

\title{Four-dimensional reflection groups and electrostatics}

\author{Maxim Olshanii         \and
        Yuri Styrkas \and
        Dmitry Yampolsky \and
        Vanja Dunjko \and
        Steven G.\ Jackson
}

\institute{M. Olshanii \at
              Department of Physics, University of Massachusetts Boston, Boston, MA 02125, USA\\
              \email{maxim.olchanyi@umb.edu}          
           \and
           Y. Styrkas \at
              Princeton High School, Princeton, NJ 08540, USA\\
              Department of Physics, University of Massachusetts Boston, Boston, MA 02125, USA
         \and
         D. Yampolsky \at
         Department of Physics, University of Massachusetts Boston, Boston, MA 02125, USA
         \and
          V. Dunjko    \at
          Department of Physics, University of Massachusetts Boston, Boston, MA 02125, USA
          \and
          S. G.\ Jackson \at
          Department of Mathematics, University of Massachusetts Boston, Boston MA 02125, USA
}


\date{\today}

\maketitle

\begin{abstract}
We present a new class of electrostatics problems that are exactly solvable by adding finitely many image charges. Given a charge at some location inside a cavity bounded by up to four conducting grounded segments of spheres: if the spheres have a symmetry derived via a stereographic projection from a 4D finite reflection group, 
then this is a solvable generalization of the familiar problem of a charge inside a spherical cavity. There are 19 three-parametric families of finite groups formed by inversions relative to at most four spheres, each member of each family giving a solvable problem. We solve a sample problem which derives from the reflection group $\gr{D}{4}$ and requires 191 image charges.
\end{abstract}

\section{Introduction}

In electrostatics, an important class of exactly solvable problems are those that can be solved by the method of images \cite{landauEandMinmedia,jackson_E_and_M}. We will restrict ourselves to cases where only finitely many images are needed and where the boundary conditions correspond to grounded conductors. So far, to the best of our knowledge \cite{Palaniappan_2011_107,Garg_2012,Smythe_1989,Jeans_2009}, the list of problems known to belong to this category has been essentially unchanged since at least Maxwell's treatise \cite{Maxwell_1873}, where it appears as follows (Art.~167, p.~206): 
\begin{quote}
The method of electrical images may be applied to any space bounded 
by plane or spherical surfaces all of which cut one another in angles which are submultiples of two right angles. In order that such a system of spherical surfaces may exist, every solid angle of the figure must be trihedral, and two of its angles must be right angles, and the third either a right angle or a submultiple of two right angles. Hence the cases in which the number of images is finite are---(1) A single spherical surface or a plane. (2) Two planes, a sphere and a plane, or two spheres intersecting at an angle $\pi/n$. (3) These two surfaces with a third, which may be either plane or spherical, cutting both orthogonally. (4) These three surfaces with a fourth cutting the first two orthogonally and the third at an angle $\pi/n'$. Of these four surfaces at least one must be spherical.
\end{quote}
According to \cite{Garg_2012} (p.~302), some of the  best-know expositions include textbooks by Jeans \cite{Jeans_2009} and Smythe \cite{Smythe_1989}. Remarkably, we were unable to find any post-Maxwell sources that treat cases (3) and (4), with the sole exception of Jeans, who only mentions the case of three planes all intersecting at right angles (Art.~211, p.~188).
  
In the present work, we will enlarge Maxwell's list. Our workhorse is the well-developed mathematical theory of \emph{finite reflection groups} \cite{Humphreys_1994}. In our context, we will need only a few basic facts from this theory to be able to successfully use it. Since the theory is perhaps unfamiliar to many readers, we will explain the main facts in detail. Reflection groups also underlie the exact solvability, via so-called Bethe ansatz, of certain models in many-body 1D quantum mechanics \cite{Olshanii_2015_105005,gaudin1983_book_english,sutherland2004_book}.

We will show that the method of images can be used to find the field induced by a point charge (placed at an arbitrary  location inside a grounded conducting cavity) if the shape of the cavity has the following properties. First, the walls of the cavity should be segments of up to four spheres (which may have infinite radius, i.e.\ can be planes). Second, the spheres in question should be (and we will devote a good portion of the text explaining what all of this means) 4D stereographic projections of great hypercircles that lie on the planes that are the generating mirrors of a 4D finite reflection group. The locations and values of the image charges are then derived, in effect, by a repeated application of the familiar formula used to find the image charge that solves the problem of a point charge next to a conducting sphere. As we will see, it is the properties of reflection groups that guarantee that there will be only finitely many image charges and that the charge values can be unambiguously assigned.

This construction includes as special cases many (but not all) elements from Maxwell's list. Therefore we will start with these familiar cases and use them to introduce the ideas that will be needed in our main construction.

\section{Basic examples}

We assume the reader is familiar with the two simplest examples of the method of images, which require just one image charge: a point charge next to a grounded conducting plane, and a point charge either inside or outside a grounded conducting sphere. In the first example, the location of the image charge is the \emph{mirror reflection} (with respect to the conducting plane) of the original charge; in the second, it is the \emph{spherical inversion} (relative to the conducting sphere) of the original. In this section, we will consider a single point charge next to a \emph{system} of planar conductors (we will return to spherical inversions in Sec.~\ref{sec_sphere_inversion}). The key question will be under what circumstances we get \emph{finitely} many images when we start reflecting the original charge relative to these planes, and then iteratively continue reflecting the images we obtain. All mathematical facts will come from Ref.~\cite{Humphreys_1994}.
 
When a line or a plane (or, in higher dimensions, a hyperplane) is used to reflect objects, we will refer to it as a `mirror'. It will be convenient to identify the mirror by a unit vector $\vu*{\alpha}$ that is normal to it. 
Thus, when we reflect a vector $\vva{\lambda}$ with respect to the plane characterized by $\vu*{\alpha}$, the result
is given by $\vva{\lambda}-2\,(\vva{\lambda}\cdot\vu*{\alpha})\,\vu*{\alpha}$. This same formula holds in all dimensions.

\begin{figure}
\centering
\includegraphics[width=0.6\columnwidth]{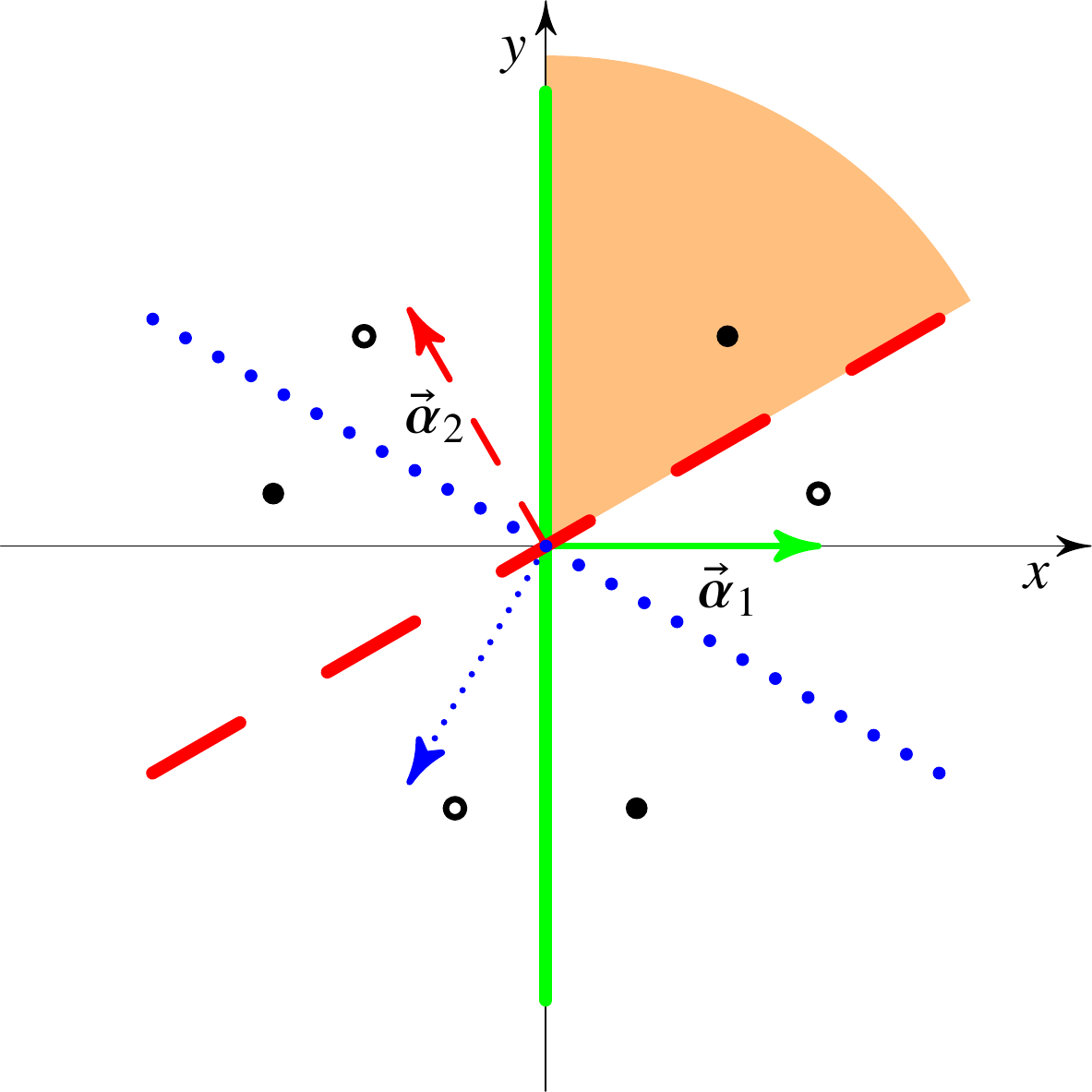} 
\caption{Two planes intersecting at an angle of $\pi/3$ and the associated system of images. In the language of finite reflection groups, this is a group that belongs to two families and can be equivalently designated as either $\gr{A}{2}$ or $\gr{I}{2}(3)$. The  solid green and dashed red lines are the \emph{generating mirrors} of the group. The vectors $\vva{\alpha}_{1}$ and $\vva{\alpha}_{2}$ are the corresponding normal vectors, whose orientation is chosen so that they constitute a system of \emph{simple roots} this reflection group. The orange wedge is the \emph{principal chamber} of the group, which is where the original charge is located, denoted by a solid black dot; the final solution will give the electric field in this wedge, assuming that the portions of the red and green lines that bound it are grounded conductors. The group has another mirror, in dotted blue, whose action can be replicated by sequences of reflections through the two generating mirrors. The set of all mirrors partitions the space into six chambers, where six is the order of the reflection group. There is one image charge in each non-principal chamber. An image charge is either the same as the original or the negative of it, and is respectively denoted either by a solid dot or by an open circle.
\label{A2_example}
}
\end{figure}

Now let's consider an example of case (2) from Maxwell's list: two planes intersecting at an angle $\pi/n$, with $n=3$; see Fig~\ref{A2_example}. 
Our region of interest is the orange wedge, which is bounded by two conducting half-planes. A charge (solid dot) is placed in this wedge, at some arbitrary location $\vva{r}$. The full planes that contain the conducting half-planes will be called the \emph{generating mirrors}, shown in solid green and dashed red in the Figure. The normal vectors  $\vva{\alpha}_{1}$ and $\vva{\alpha}_{2}$ are chosen so that a vector $\vva{r}$ (whose initial point is the origin) lies in the orange wedge if and only if the projections $\vva{r}\cdot\vva{\alpha}_{1}$ and $\vva{r}\cdot\vva{\alpha}_{2}$ are both positive. When they are chosen this way, they are called \emph{simple roots} of the reflection group.

Now we start reflecting $\vva{r}$ with respect to the generating mirrors. More precisely, let $S$ be a set in which we will be collecting distinct outcomes of reflections of the initial vector $\vva{r}$; we initialize $S$ so that it contains just this initial vector. Now we iterate the following process: for each element of $S$, we reflect that element once through each generating mirror, adding the outcome to $S$ unless it is already there---and then repeat with the new, enlarged $S$. In the present case, we will find that once we generate the five image charges, so that $S$ contains six charges total, the process stops: a reflection of any one of these charges about either generating mirror produces one of the charges that is already in $S$. One can prove that if the angle between the generating mirrors were something other than $\pi/n$, with $n$ a positive integer, the process would never stop: there would be an infinity of image charges. 

Note that there is a third plane, shown in dotted blue, that we can also be considered to be a mirror. Its location be obtained by reflecting one of the two generating mirrors about the other, and its action is equivalent to a sequence of reflections about the generating mirrors; in other words, adding this extra mirror does not result in any extra image charges. However, it is useful to consider the full system of mirrors, because it leads to the concept of \emph{chambers}. Note that the three mirrors in the Figure partition the space into six congruent regions, which are the chambers. It turns out there is one charge (image or original) per chamber. This total number of charges (which is the total number of chambers), here six, is the \emph{order} of the reflection group. 

There are final two facts to notice. First, not every sequence of reflections about the mirrors is a reflection about some (other) mirror. Consider, for example, the solid dot in the lower right chamber bounded by the solid green and dotted blue mirrors. It cannot be obtained from the original charge by any single reflection; in fact, what links these two is a rotation by $2\pi/3$. Second, all the charges are at the same distance from the origin, and so lie on a circle.

This procedure can be generalized to any number of dimensions, and to any number of mirrors. Pick some number of mirrors, and a vector $\vva{r}$. Initialize a set $S$ as containing just $\vva{r}$. Iterate the following: reflect each element of $S$ with respect to each of the chosen mirrors, adding the outcome to $S$ unless it is already there; repeat with the enlarged $S$. For almost all choices of the mirrors, the size of $S$ will grow without bound for almost all choices of $\vva{r}$. The cases where it does not, where we only get a finite number of images no matter how $\vva{r}$ is picked, are special: we say that the corresponding system of mirrors generates a \emph{finite reflection group}, whose order is the maximal size $S$ can have no matter how $\vva{r}$ is picked. (In fact, a generic choice of $\vva{r}$ results in the maximal size of $S$). The reflection group may have extra mirrors, which can be obtained by initializing $S$ with our chosen mirrors, and iterating as before; we will call the final result the \emph{full set} (or system) of mirrors for the group. The full set of mirrors partitions all of space into disjoint regions called \emph{chambers}. The original vector $\vva{r}$ is in one chamber, and all the other chambers contain exactly one image; thus, there are as many images as there are chambers, and their number is called the \emph{order} of the group.

When the originally chosen mirrors do generate a finite reflection group, it may be possible to discard some of the originally chosen mirrors and yet have the iterative procedure (using just the retained mirrors) still generate the same images as before. More generally, we may start with the full set of mirrors, and do the following procedure. We start discarding the mirrors one by one; whenever we discard one, we test to see if the remaining ones still generate all the images. If not, we reinstate the mirror; if yes, we leave it out. Then we continue discarding mirrors, but any mirror we had to reinstate at any previous step we don't attempt to discard again. Eventually we will end up with only mirrors that we had to reinstate at some point. It turns out that no matter in what order we were attempting to discard the mirrors, the final set of mirrors will always have the same size, called the \emph{rank} of the reflection group; we will denote it by \rnk.
Any such minimal set of \rnk mirrors, which generates the full group, is a system of \emph{generating mirrors} for the group.  All other mirrors can be obtained by initializing $S$ with just the generating mirrors, and iterating as before. The size of the full system of mirrors is generally smaller than the order of the group (the number of chambers), often vastly so. 

For any system of generating mirrors, one chamber will be enclosed entirely by the generating mirrors in that set; this is the \emph{principal chamber}. (In some cases there could be more than one such chamber; in that case, just pick one.) The \rnk normal vectors $\vva{\alpha}_{i}$, $i=1,\,\ldots,\,\rnk$ that characterize the generating mirrors can be chosen so that a vector $\vva{r}$ belongs to the principal chamber if and only if $\vva{r}\cdot\vva{\alpha}_{i}>0$ for all $i=1,\,\ldots,\,\rnk$. In that case, the set of vectors $\vva{\alpha}_{i}$ is called a set of \emph{simple roots} of the group, usually denoted by $\Delta$. It is always a linearly independent set, but generally not an orthogonal one. Thus, the rank is the lowest dimension that a space must have in order for the reflection group to be realizable in it, and the simple roots are a basis for it (though not an orthonormal one). Since we will ultimately be dealing with the 4D space, we can use any reflection group of rank 4 or lower. It turns out there are 19 such groups, and we list all of them in Tables~\ref{reflection_groups_1} and \ref{reflection_groups_2}. For each, we give the Coxeter diagram (a concept we will explain below), the number of mirrors, the order of the group, and the simple roots. 

\newcounter{itm}
\newcommand{\ittm}{\roman{itm}\protect\stepcounter{itm}}
\newcommand{\cosm}[1]{c_{#1}}
\begingroup
\centering
\begin{table}
\centering
  \includegraphics[width=0.8\columnwidth]{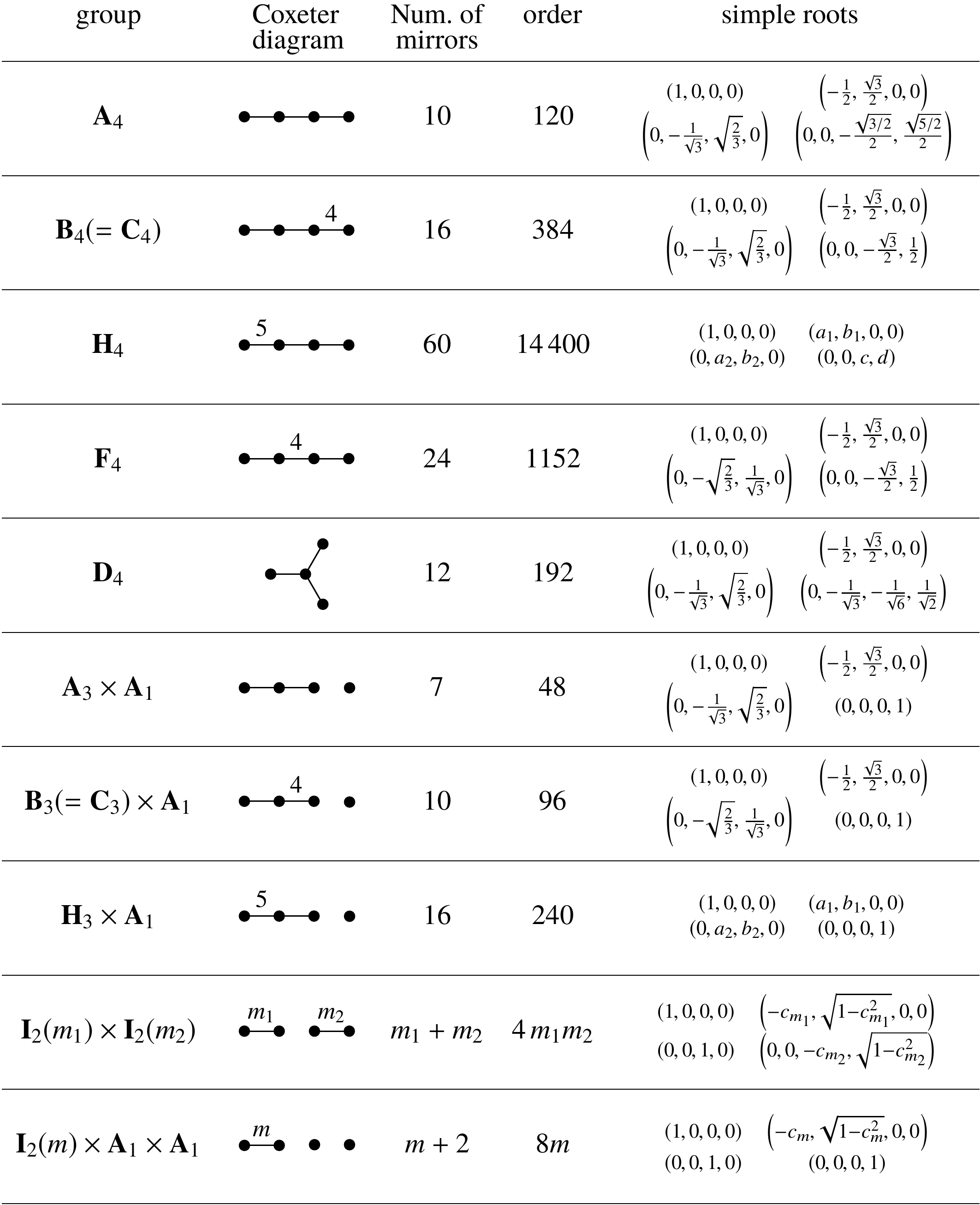}
  \caption{The 19 finite reflection groups of rank four or less, part 1 of the list. The subscript to the letter name of the group denotes the rank. Thus, the rank of the group is the sum of the subscripts. For example, both \gr{H}{3} and $\gr{I}{2}(m)\times\gr{A}{1}$ have rank 3, and so they each have three simple roots.   Here $m_{1},\,m_{2},\,m=3,\,4,\,\ldots\,.$ In groups that contain $\gr{I}{2}(m)$, $\cosm{m}=\cos\frac{\pi}{m}$. In \gr{H}{4} and groups that contain \gr{H}{3}, $a_{1}={\scriptstyle -\frac{1}{4} \left(1+\sqrt{5}\right)}$, $b_{1}={\scriptstyle \frac{1}{2} \sqrt{\frac{1}{2} \left(5-\sqrt{5}\right)}}$, $a_{2}={\scriptstyle -\sqrt{\frac{1}{10} \left(5+\sqrt{5}\right)}}$, $b_{2}={\scriptstyle \frac{1}{2} \sqrt{1+\frac{2}{\sqrt{5}}} \left(3-\sqrt{5}\right)}$; further, in \gr{H}{4}, $c={\scriptstyle \frac{1}{4} \sqrt{1+\frac{2}{\sqrt{5}}} \left(\sqrt{5}-5\right)}$, $d={\scriptstyle \frac{1}{2} \sqrt{\frac{1}{2} \left(3-\sqrt{5}\right)}}$. Note the following:
{\protect\setcounter{itm}{1}}
\ittm. \gr{A}{2}, \gr{B}{2}, \gr{H}{2}, and \gr{G}{2} are included respectively as $\gr{I}{2}(3)$ through $\gr{I}{2}(6)$; \ittm. \gr{D}{n} only exist for $n\geqslant 4$; \ittm. \gr{F}{n} only exists for $n=4$; and, \ittm. \gr{H}{n} only exist for $n=2,\,3,\,4$.  
  }
\label{reflection_groups_1}
\end{table}
\begin{table}
\centering
  \includegraphics[width=0.8\columnwidth]{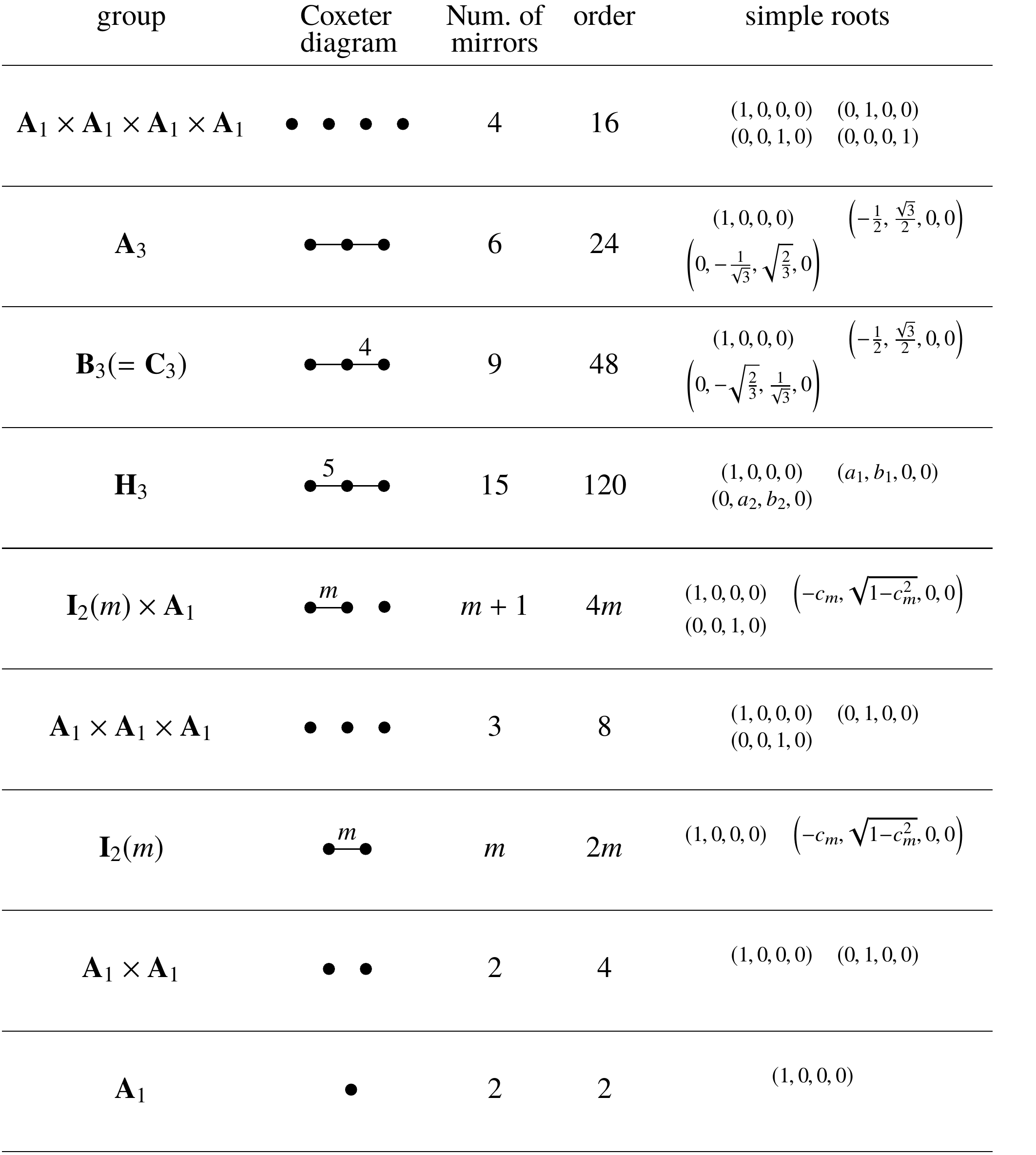}
  \caption{The 19 finite reflection groups of rank four or less, part 2 of the list. As before, in groups that contain $\gr{I}{2}(m)$, $\cosm{m}=\cos\frac{\pi}{m}$. The final entry in the table above corresponds to the classic problem with one spherical cavity---the very inspiration for this paper. 
  }
\label{reflection_groups_2}
\end{table}
\endgroup
%
%
Lastly, if one considers all the mirrors, one can have two normal vectors (which are negatives of each other) per mirror. These vectors are called the \emph{roots}, and we see that there are twice as many roots as there are mirrors.

Note that if a system of mirrors of a group is rotated as a whole by any amount, the result still constitutes the same finite reflection group. This property will be important below.
%

All finite reflection groups, in all spatial dimensions, have been classified and their properties are very well understood. As is clear from the previous discussion, each reflection group is completely defined by its simple roots. In particular, there are standard choices for the simple roots of all possible reflection groups. They can be looked up in any number of sources, and in particular in Ref.~\cite{Humphreys_1994}. We also give them in Tables~\ref{reflection_groups_1} and \ref{reflection_groups_2}, but we should note that the way we wrote them is not the standard choice for their presentation. Our motivation for this will become clear below.

The reflection groups corresponding to two planes meeting at an angle of $\pi/n$ are called $\gr{I}{2}(n)$. Four of them have aliases, because they also count as members of other families of groups; see the caption of Table~\ref{reflection_groups_1}. In the group notation, the subscript always denotes the rank. Note that the mirrors of $\gr{I}{2}(n)$ are the planes of symmetry of a regular $n$-gon.

\section{First extension of Maxwell's list of solvable problems: irreducible finite reflection groups of rank 3}
\label{3D_planar_problems}

When we consider the problem in Fig.~\ref{A2_example} (or indeed any problem where two planes meet at an angle $\pi/n$), it is clear that the problem remains solvable if we add another conducting plane parallel to the $xy$-plane: all we have to do is reflect all the charges about that plane and invert their signs. The resulting system of planes is also a reflection group, denoted by $\gr{I}{2}(n) \times \gr{A}{1}$. Here $\gr{A}{1}$ is the usual notation for the rank-1 group consisting of a single mirror, and `$\times$' denotes the \emph{direct product} of groups. The latter is easiest to understand by considering how to construct the mirrors and the simple roots of the direct product of two groups. 

\subsection{Direct products of reflection groups; reducible and irreducible groups}

Let $g_{1}$ be a finite reflection group  of rank $r_{1}$, order $o_{1}$, and with a total of $m_{1}$ mirrors, and let $g_{2}$ be a reflection group whose corresponding numbers are $r_{2}$, $o_{2}$, and $m_{2}$. To construct the system of mirrors, simple roots, and chambers for $g_{1} \times g_{2}$, we will need a space of at least $r_{1}+r_{2}$ dimensions; for definiteness, let's choose $\mathbb{R}^{r_{1}+r_{2}}$. The simple roots of $g_{1}$ define an $r_{1}$-dimensional hyperplane in that space, while the simple roots of $g_{2}$ can always be chosen in $\mathbb{R}^{r_{1}+r_{2}}$ so that they are all orthogonal to all simple roots of $g_{1}$. An obvious way to do this is to take the simple roots of $g_{1}$ (which can always be taken as $r_{1}$-dimensional row vectors) and append $r_{2}$ zeros to these row vectors; then we similarly \emph{pre-}pend $r_{1}$ zeros to the $r_{2}$-dimensional row vectors that are the simple roots of $g_{2}$. Once such a choice is made, collectively the simple roots of $g_{1}$ and $g_{2}$ will be the simple roots of $g_{1} \times g_{2}$, which thus has rank $r_{1}+r_{2}$, and whose total number of mirrors is $m_{1}+m_{2}$. The order of the direct product turns out to be the product of the orders, $o_{1} o_{2}$. A similar procedure can be used for direct products of more than two groups. 

In the example of $\gr{I}{2}(n) \times \gr{A}{1}$, the `hyperplane' was chosen to be the $xy$-plane. Group $\gr{I}{2}(n)$ is of rank 2 with simple roots $(1,\,0)$ and $(-1/2,\,\sqrt{3}/2)$, while $\gr{A}{1}$ is of rank 1 with simple root  $(1)$. So the rank of  $\gr{I}{2}(n) \times \gr{A}{1}$ is $2+1=3$. It is of order $4n$ and has a total of $n+1$ mirrors. We obtain its simple roots by appending $3-2=1$ zeros to the simple roots of $\gr{I}{2}(n)$, obtaining $(1,\,0,\,0)$ and $(-1/2,\,\sqrt{3}/2,\,0)$, and by prepending $3-1=2$ zeros to the simple root of $\gr{A}{1}$, obtaining $(0,\,0,\,1)$.

A a group is called \emph{irreducible} if it cannot be written as a direct product of two other groups; if it can, then it is \emph{reducible}. In Tables~\ref{reflection_groups_1} and \ref{reflection_groups_2}, all the groups that are \emph{not} written as direct products are in fact irreducible.

\subsection{First instances of new solvable problems: systems related to irreducible groups of rank 3} 

Considering how we solved the electrostatics problem associated with $\gr{I}{2}(n) \times \gr{A}{1}$, we might suspect that all finite reflection groups of rank 3 will generate a similar solvable electrostatics problem for a point charge in a cavity bounded by three grounded conducting planar walls. This is in fact correct, as we will show momentarily. Looking at Tables~\ref{reflection_groups_1} and \ref{reflection_groups_2}, we see that, besides the reducible groups of rank 3 (namely, $\gr{A}{1} \times  \gr{A}{1} \times \gr{A}{1}$ and the family $\gr{I}{2}(n) \times \gr{A}{1}$), there are also three irreducible ones: $\gr{A}{3}$, $\gr{B}{3}$ (alias $\gr{C}{3}$), and $\gr{H}{3}$. It is these that give solvable problems that were not on Maxwell's list. To construct the generating mirrors in 3D, we need the 3D simple roots of these groups. These can be obtained from Table~\ref{reflection_groups_2}. True, they are given there as 4D vectors, but we actually originally obtained them as 3D vectors, and then embedded them in 4D space by appending a zero as the fourth coordinate. Thus, now we are just reversing this, discarding the zero fourth coordinate.


We should emphasize that all of the solvable problems we are presently discussing, which are related to reflection groups in 3D, can also be obtained as special cases of our general 4D procedure, which we will be explaining shortly. Nevertheless, it may be helpful to outline how one constructs a solvable problem in this simpler 3D case, where we don't use a stereographic projection. We will then proceed to use the 3D case to illustrate the stereographic projection in the 3D setting, which is of course easier to visualize than the 4D setting we will eventually be using.

\subsection{An example: group \gr{A}{3}}  

For definiteness, let's pick the group $\gr{A}{3}$, which is the symmetry group of the regular tetrahedron. For the groups $\gr{A}{n}$, the standard form of the simple roots is given as embedded in the space of $n+1$ dimensions, where they define an $n$-dimensional hyperplane. The three simple roots of $\gr{A}{3}$ are given in the literature as $(1,\,-1,\,0,\,0)$, $(0,\,1,\,-1,\,0)$, and $(0,\,0,\,1,\,-1)$. It is clear that none of the simple roots is orthogonal to both of the other ones, so this case indeed is not on Maxwell's list. To find what these vectors are in 3D space, we apply Gram-Schmidt process to them to generate an orthonormal basis for the 3D space that they span, and then find their coordinates relative to that basis. The result (after normalization) is $\vva{\alpha}_{1}=(1,\,0,\,0)$, $\vva{\alpha}_{2}=(-1/2,\,\sqrt{3}/2,\,0)$, and $\vva{\alpha}_{3}=(0,\,-1/\sqrt{3},\,\sqrt{2/3})$.  This is exactly what appears in Table~\ref{reflection_groups_2}, after we discard the fourth (zero) coordinate.

\begin{figure}
\centering
\includegraphics[width=0.8\columnwidth]{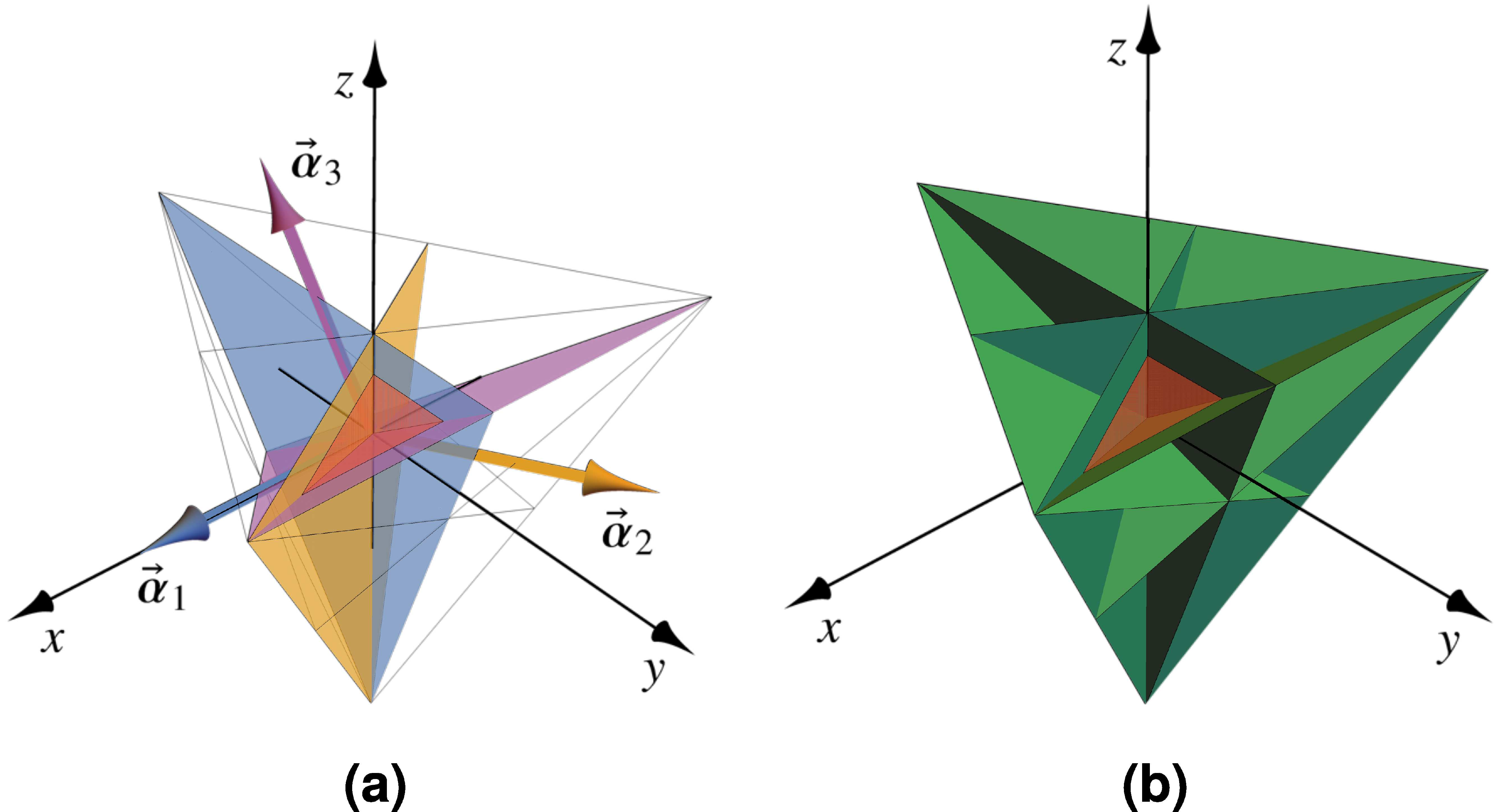}
 \caption{Geometry of \gr{A}{3}. {\sffamily \textbf{(a)}} The simple roots and the corresponding generating mirrors. A part of the principal chamber is shown in red. {\sffamily \textbf{(b)}} The full system of mirrors, with part of the principal chamber in red.
 \label{A3_example}}
\end{figure}

As we said, this group is a symmetry group of the regular tetrahedron. We can start from the three simple roots we've just written down; see Fig.~\ref{A3_example}(a).
They define three generating mirrors. We can iteratively (as we described above) reflect the generating mirrors about each other until no new mirrors are produced. In this case, we will get three extra mirrors, for a total of six. They are depicted in Fig.~\ref{A3_example}(b). The six mirrors will be the six planes of symmetry of the regular tetrahedron, and so may be described as follows. Consider a tetrahedron. Pick a pair of vertices, and consider the edge that connects them. Now look at the remaining pair of vertices, and consider the midpoint of the edge that connects \emph{them}. The former edge and the latter midpoint define a plane. It should be clear from the construction that this is a plane of symmetry of the tetrahedron. There are $\binom{4}{2}=6$ ways to pick a pair of vertices, and thus there are six planes of symmetry. Any three can be taken as generating mirrors, and the three simple roots we've written down will indeed give a triplet of such planes. All the mirrors meet at a point (this is the case with all finite reflection groups, in all dimensions), and we will set the origin of our coordinate system at that point. Note that we still have the freedom to rotate our system of mirrors around the origin. The full set of mirrors partition the space into chambers, which are visible in Fig.~\ref{A3_example}(b). To determine their number, note that a tetrahedron has 4 faces, and on each face we see into six chambers. So there are 24 chambers total, which is also the order of $\gr{A}{3}$. One of the 24 chambers is the principal chamber, defined as the set of points $\vva{r}$ such that all of $\vva{r}\cdot\vva{\alpha}_{i}$ is positive for $i=1,\,2,\,3$. In Fig.~\ref{A3_example}(a) and (b), a part of the principal chamber is shown in red. All other chambers are the images of the principal chamber `under the action' of the reflection group. This means that if we start iteratively reflecting the entire principal chamber about the generating mirrors, we will eventually generate the remaining 23 chambers, and after that we will generate no further ones. 

\subsection{Constructing a solvable electrostatics problem}
\label{solvable_planar_problem}

To construct a solvable electrostatic problem, consider the principal chamber. It is enclosed by parts of the generating mirrors. Suppose these parts are grounded conducting surfaces. Place a point charge inside the principal chamber: our electrostatics problem is to find the electric field inside this chamber. To solve it by the method of images, iteratively reflect that point charge through the generating mirrors until 23 distinctive images are created, one in each non-principal chamber; further reflections will not generate any new ones. For future reference, note that all charges are the same distance from the origin, and so lie on the surface of a sphere. The magnitudes of all the charges are the same; the only nonobvious question is whether it is possible to assign the charge \emph{signs} in a consistent manner. What is required is that any pair of charges related by a single reflection have opposite signs. That way, their contribution to the potential at the mirror between them will be zero. Furthermore, our iterative construction guarantees that for every charge and every mirror, there is a charge that is the reflection of the first charge relative to that mirror. Thus, at every mirror, the potentials due to all the charges will cancel out.  

Here is how to assign the charge signs. Note that every image charge is the result of some sequence of reflections from the original charge. Let the image charge have the same sign as the original if it takes an even number of reflections to generate it from the original, and let it have the opposite sign if it takes an odd number. This guarantees any pair of charges related by a single reflection have opposite signs: after all, they will be reachable from the original charge through reflection sequences whose lengths differ by one. Thus the lengths will have opposite parities, and so the image charges will have opposite signs.
 
The only possible problem is this: note that each image charge is reachable from the original one through a variety of reflection sequences, which may well have unequal lengths. And if some of these lengths are odd numbers while others are even, we are in trouble. Luckily, a general theorem of reflection groups guarantees that the parity of the length will in fact always be the same. Thus the values of the image charges can be unambiguously assigned, and the full system of charges will solve our electrostatics problem in the principal chamber.

\subsection{Coxeter diagrams} 
At this point we can introduce \emph{Coxeter diagrams}, which are a basic tool in the theory of finite reflection groups. In brief: since the relative orientation of simple roots is all that matters, simple roots (and thus the finite reflection group as a whole) can be efficiently represented in a diagrammatic form, as follows. Each simple root is represented by a node. In most finite reflections groups, an overwhelming majority of simple roots are mutually orthogonal, and so we connect two nodes by an edge only if the corresponding simple roots are \emph{not} mutually orthogonal. Each edge therefore corresponds to two mirrors intersecting at a dihedral angle which is not $\pi/2$. The angle is, however, always of the form $\pi/m$, with $m$ an integer. We place $m$ above the edge, signifying that the angle between the mirrors is $\pi/m$. It turns out that $m=3$ is by far the most common label, so by convention we leave the edge unlabeled if $m=3$ (whereas we omit drawing the edge altogether if $m=2$). The result is called a \emph{Coxeter diagram}.

Using simple roots, the label $m$ (which we place above the edge connecting the nodes that represent two simple roots $\vva{\alpha}_{i}$ and $\vva{\alpha}_{j}$) is given by $m=\pi/\arccos \left(-\vva{\alpha}_{i}\cdot\vva{\alpha}_{j}\right)$, where we assumed the simple roots are normalized. 

A group is irreducible just in case its Coxeter diagram is connected. It should now be easy to verify that these are the Coxeter diagrams corresponding to the groups whose simple roots we've already written down: for $\gr{A}{1}$, 
\dynkin[Coxeter]{A}{1}\,; 
for  $\gr{I}{2}(m)$, 
\dynkin[Coxeter,gonality=m]{I}{2}\,; 
for $\gr{I}{2}(n) \times \gr{A}{1}$,  
\dynkin[Coxeter,gonality=m]{I}{2}\,\dynkin[Coxeter]{A}{1}\,; 
and for $\gr{A}{3}$, 
\dynkin[Coxeter]{A}{3}\,. 
For example, note the inner products of the simple roots of \gr{A}{3}: $\vva{\alpha}_{1}\cdot\vva{\alpha}_{2}=\vva{\alpha}_{2}\cdot\vva{\alpha}_{3}=-1/2$ and $\vva{\alpha}_{1}\cdot\vva{\alpha}_{3}=0$. This corresponds to $m=3,\,3,\,2$. We see that there are indeed two unlabeled edges corresponding to the two instances of $m=3$, and that the first and last nodes are not connected by an edge, corresponding to $m=2$.
%

\section{Stereographic projection}

Our main result concerns reflection groups in 4D, and the way we will go from the abstract 4D space to our physical 3D one will be via a stereographic projection. Here we will first describe our procedure in the more familiar 3D setting. For an account of the stereographic projection from 3D to 2D, we refer to Ref.~\cite{Rosenfeld_1977}, which contains basic facts and proofs using Eucledian geometry. 

\begin{figure}
\centering
\includegraphics[width=0.85\columnwidth]{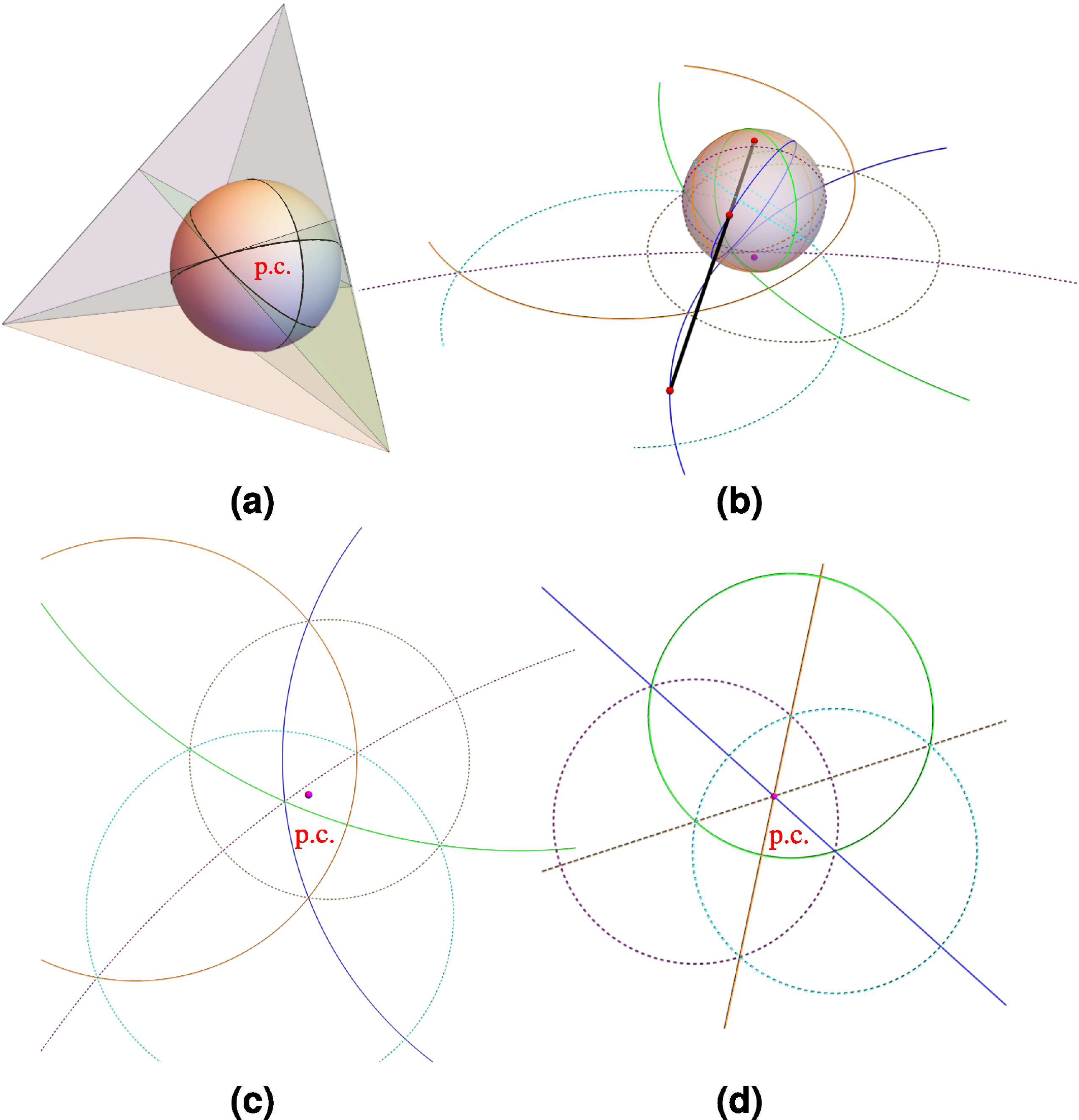} 
\caption{Stereographic projection of reflection group geometry from 3D to 2D, on the example of \gr{A}{3}. The regions corresponding to the principal chamber are marked by red letters `p.c.' {\sffamily \textbf{(a)}} The great circles on a sphere that are the intersections of the sphere and the generating mirrors. {\sffamily \textbf{(b)}} The stereographic projection of the great circles from (a) onto a plane tangent to the sphere at the sphere's `South Pole', which is marked by the point in magenta. The solid lines correspond the generating mirrors, while the dotted ones correspond to the other mirrors {\sffamily \textbf{(c)}} The resulting 2D system of circles. The `South Pole' is marked by the point in magenta. The region corresponding to the principal chamber is bounded by circular arcs. {\sffamily \textbf{(d)}} If the system of mirrors were tilted so that two of the generating mirrors passed through the South Pole, then two of solid arcs become straight lines. 
 \label{tetrahedron}
 }
\end{figure}

Choose a system of generating mirrors in 3D, e.g.\ corresponding to \gr{A}{3}. The generating mirrors all meet at a point. Choose a sphere whose center is at that point, of any radius $R'$. Form the intersections of this sphere and the generating mirrors. The result is three great circles, as shown in Fig.~\ref{tetrahedron}(a). 
The region corresponding to the principal chamber is a spherical triangle. Now choose some point on the sphere to be the `South Pole'; the point diametrically opposite is then the `North Pole'. Consider the 2D plane $S$ tangent to the sphere at the South Pole. The stereographic projection of points on the sphere to this plane is then found as follows. Pick a point of interest, $P'$,  on the sphere. Consider the line $L$ that passes through both the North Pole and the point of interest. The stereographic projection of the point of interest is the point $P$ on the tangent plane $S$ at which the line $L$ crosses the that plane. This is shown in Fig.~\ref{tetrahedron}(b). The `North Pole' is marked by the red point on the topmost end of the black line. This line shows the projection of a point on the blue great circle. Note that the system of mirrors was tilted so that the spherical triangle corresponding to the principal chamber is near the South Pole, but also so that no great circle passes through the South Pole. The resulting projection is shown in (c). If the system of mirrors were tilted so that two of the generating mirrors passed through the South Pole, then two of the arcs become straight lines. The third one still must be a circular ark, because the three great circles corresponding to the generating mirrors never all meet at one point.

If we pick a coordinate system such that its origin is at the center of the sphere, so that the South Pole is at $(0,\,0,\,-R')$, it is easy to show that the stereographic projection $\left(x,\,y\right)$ of a point $(x',\,y',\,z')$ on the sphere (i.e.\ such that $(x')^{2}+(y')^{2}+(z')^{2}=(R')^{2}$) is given by
\[
 \left(x,\,y\right) = \frac{2R'}{R'-z'}\left(x',\,y'\right)\,.
\]
If the point on the sphere has spherical coordinates $(R',\,\theta',\,\phi')$ so that $(x',\,y',\,z')=R'\, (\sin \theta' \cos \phi',\,\sin \theta' \sin \phi',\,\cos \theta')$, then the polar coordinates $(\rho,\,\phi)$ of the projected point are $\rho=2R'\left|\cot\frac{\theta'}{2}\right|$ and $\phi=\phi'$ (and of course the $z$-coordinate, if we choose to include it, is $-R'$). Relevant properties of the stereographic projections are as follows. 

First, a great circle on the sphere that passes through the poles is mapped to a line in the plane (but it will be convenient to consider this a limiting case of a circle of as its radius goes to infinity). If the great circle does not pass through the poles and lies in the plane whose unit normal is $\bm{n}=(n_{x},\,n_{y},\,n_{z})=(\sin \theta_{\bm{n}} \cos \phi_{\bm{n}},\,\sin \theta_{\bm{n}} \sin \phi_{\bm{n}},\,\cos \theta_{\bm{n}})$, then it is mapped to a circle $C$ in the plane, whose radius is $R=2 R' \left|\sec \theta_{\bm{n}}\right|$ and whose center is $\vva{O}=-2 R' (\sec \theta_{\bm{n}})\,(n_{x},\,n_{y},0)+(0,\,0,\,-R')$. 

Second, recall that the great circle is an intersection of a mirror and the chosen sphere. Pick a point $\vva{p}'_{1}$ on the sphere and reflect it through this mirror. The reflection $\vva{p}'_{2}$ is also on the surface of the sphere. Now consider the stereographic projections $\vva{p}_{1}$ and $\vva{p}_{2}$ of the two points. It turns out that one of the points is inside the circle $C$, one outside, and that they are \emph{circle inversions} of each other, with $C$ as the inversion circle. This means that $\vva{p}_{1}$, $\vva{O}$, and $\vva{p}_{2}$ lie on a line, and that the distances from the points to the center of the circle satisfy $d(\vva{p}_{1},\,\vva{O})\,d(\vva{p}_{2},\,\vva{O})=R^{2}$, where $d(\vva{p},\vva{O})=\nrm{\vva{p}-\vva{O}}$. Here for any vector vector $\vva{v}$, $\nrm{\vva{v}}$ is its norm, defined as $\nrm{\vva{v}}=\sqrt{\vva{v}\cdot\vva{v}}$.

Third, stereographic projection is \emph{conformal}, meaning that if two curves on the surface of the sphere intersect at some angle, the curves that are their projections will intersect at the same angle.

We should mention that we were unable to find an \emph{explicit} mention of the second property in the literature, except for the special case where the great circle is parallel to the plane to which we are projecting. In particular, it is not mentioned in Ref.~\cite{Rosenfeld_1977}. Our explicit proofs (for 3D$\to$2D and 4D$\to$3D cases) are, so far, brute-force computations which we will not reproduce.

\section{An outline of our procedure in 4D}
\label{4D_outline}

We will now have to repeat the procedure of the previous section with all numbers of spatial dimensions raised by one. Pick a reflection group $g$; let $\ell$ be its rank and $\left|g\right|$ its order. Place its $\ell$ generating mirrors in the 4D space. This means that $\ell$ must be 4 or less. There are 19 finite reflection groups of rank 4 or less, and they are all listed in Tables~\ref{reflection_groups_1} and \ref{reflection_groups_2}. The generating mirrors of $g$ always meet at a point, and we choose a coordinate system whose origin is at that point. We are free to rotate our system of mirrors about the origin; below, we will briefly discuss how to parametrize 4D rotations.

Now pick a radius $\mathcal{R}$ and consider a 4D hypersphere (a 3D object), centered at the origin, of that radius. The intersections of the hypersphere and the generating mirrors are great hypercircles $S'_{i}$, $i=1,\,\ldots,\,\ell$. The intersection of the sphere and the principal chamber, which we will denote $A'$, is a 3D region on the hypersphere. Choose a point $P'_{1}$ in $A$, and find all its images $P'_{j}$, $j=2,\,\ldots,\,\left|g\right|$, relative to the generating mirrors. 

Now we stereographically project to the 3D hyperplane tangent to the hypersphere at its South Pole, $(0,\,0,\,0,\,-\mathcal{R})$. The $\ell$ projections of the great hypercircles are $S_{i}$, each of which is either a plane (if it comes from a great hypercircle that passes through the poles) or else a sphere. Note that by 4D-rotating our system of mirrors we can make up to three great hypercircles pass through the poles.

The projections of all the points are $P_{i}$, $i=1,\,\ldots,\,\ell$. The projection of $A'$ is a 3D region bounded by segments of $S_{i}$'s, and it contains $P_{1}$. Note that although $P_{i}$, $i=2,\,\ldots,\,\ell$ were obtained via the stereographic projection of $P'_{i}$'s from 4D, we could obtain them in a different way, entirely within 3D. Namely, we just iteratively sphere-invert $P_{1}$ relative to the spheres $S_{i}$ (for any $S_{i}$ that is a plane, we just reflect as usual). Below, we will in fact refer to this procedure to figure out how to assign the values to the image charges. 

Note that the above implies that we can use the stereographic projection to generate a finite group of sphere inversions (possibly supplemented by plane reflections). In fact, our original motivation for this project was to find out what concept plays the same role for sphere inversions as finite reflection groups do for plane reflections; it just turned out that sphere inversions are simply related to plane reflections (in one spatial dimension higher) through the stereographic projection.

\subsection{Comparison to the Kelvin transform}
At this point, we should mention the well-known Kelvin transform \cite{Kellogg_1967}, which, given any problem solvable by images, can be used to generate a different solvable problem. Physics sources tend to refer to it as the `method of inversion' \cite{landauEandMinmedia,Jeans_2009,Smythe_1989}, but since sphere inversions appear in our procedure in a different way, we will not use that name. The Kelvin transform involves sphere-inverting the geometry of the original problem relative to an arbitrary inversion sphere, and adjusting the resulting charges in a particular way. Sphere inversion transforms planes and spheres to planes and spheres, and in particular can transform planes into spheres. Indeed, the Kelvin transform is the origin of the spherical surfaces in Maxwell's list of solvable problems. 

These properties of spherical inversions sound remarkably similar to those of stereographic projections. This is not an accident: the stereographic projection of a sphere $s$ is equivalent to a spherical inversion of $s$ relative to an inversion sphere $S$, whose radius is the diameter of $s$, centered at the North Pole of $s$. In general, under spherical inversion, a sphere that passes through the origin of the inversion sphere gets mapped to a plane.  If it is further true that the radius of $S$ is twice the radius of $s$, simple geometry shows that the spherical inversion is equivalent to a stereographic projection.  

Since our 4D procedure also generates solvable problems involving intersecting spheres and planes, one might suspect that our construction and the construction using the Kelvin transform are closely related. However, it appears that this is not the case, and that neither is reducible to the other. 

To see that the Kelvin transform construction is not more general than our 4D construction, note that Kelvin transform requires a solvable 3D problem as input. If we don't use our 4D construction, all solvable problems must ultimately come from reflection groups of rank 3 or lower (via the methods described in Sec.~\ref{3D_planar_problems}). The geometries that result from applying the Kelvin transform to those cannot be equivalent to those that are obtained (via our 4D procedure) from \emph{irreducible} reflection groups of rank 4. To see this, recall that both the spherical inversion and the stereographic projection are \emph{conformal}, i.e.\ angle-preserving. On the other hand, reflection groups are completely determined by the dihedral angles between their generating mirrors (as encoded in their Coxeter diagrams). After transformation (be it a spherical inversion or a stereographic projection), whatever spheres and planes we obtain will still cross each other at those same angles. So a spherical inversion of a 3D system of generating mirrors cannot produce the same angles as a stereographic projection of an irreducible rank-4 system.

And to see that our 4D construction is not more general than the Kelvin transform one, note that when our construction maps the principal chamber to an area completely enclosed by spherical or planar surfaces, the original charge should be placed within this area. On the other hand, the Kelvin transform can be used to produce problems where the charge is on the outside: just start with a $\pi/n$ wedge (with a point charge inside of it) and choose an inversion circle inside this wedge. After a spherical inversion, the half-planes that define the wedge will become intersecting spheres, with the appropriate parts of them missing so that the result is a single `double bubble' surface, which completely encloses a finite region of space. In this case, it is the image charges (of the original solvable-wedge problem) whose Kelvin transforms will end up on the inside, while the Kelvin transform of the original charge will be on the outside.  

In our 4D construction, after we stereographically project the great hypercircles to ordinary 3D spheres and planes, we have the locations of all the conducting surfaces and all the charges, the original one and the image charges. However, we still don't know which values we need to assign to the image charges. To determine these values, we will need to use sphere inversions. Note that these sphere inversions are not Kelvin transforms, because neither the bounding walls not the original charge are transformed. Let us describe this type of spherical inversion quantitatively.

\section{Sphere inversion for a single spherical conducting cavity}
\label{sec_sphere_inversion}
\begin{figure}[!ht]
\centering
\includegraphics[width=0.9\columnwidth]{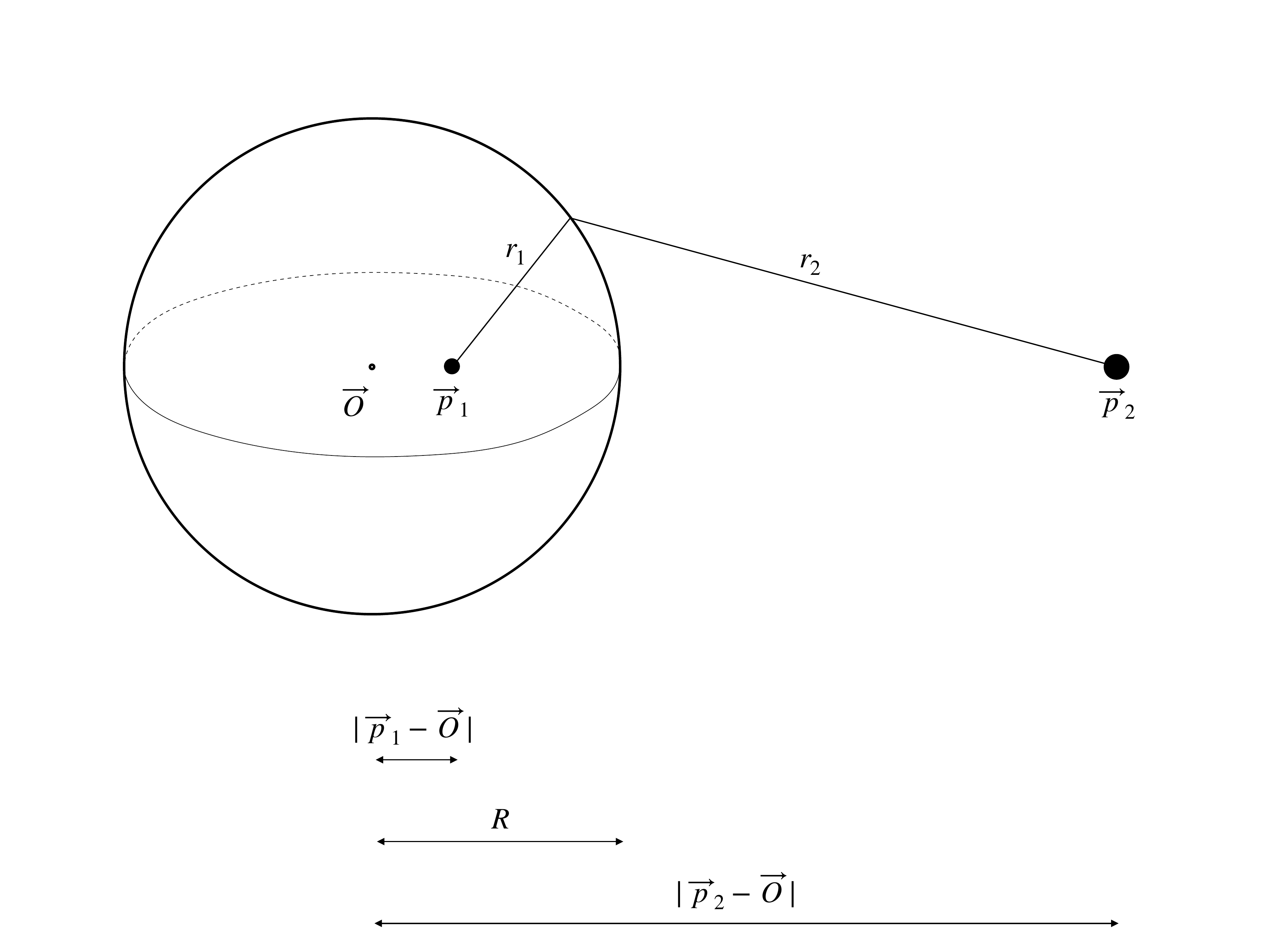}
\caption{
A charge inside a conducting spherical cavity: notations and definitions. $\bm{O}$ is the center of the cavity. The vector $\bm{p}_{1}$  is the position of the physical charge and $r_{1}$ is the distance between a particular point on the cavity wall and the physical charge, while $\bm{p}_{2}$ and $r_{2}$ are the corresponding quantities for the image charge. The distances $|\bm{p}_{1}-\bm{O}|$ and $|\bm{p}_{2}-\bm{O}|$ govern the assignment of the value of the image charge. 
             }
\label{f:basic_geometry}        
\end{figure}

Consider a point charge $q_{1}$ at a position $\bm{p}_{1}$ inside an empty spherical cavity of radius $R$ centered at a point $\bm{O}$, surrounded by a grounded conductor (Fig.~\ref{f:basic_geometry}). Consider a point $\bm{p}_{2}$ related to the point $\bm{p}_{1}$ through a sphere inversion, with the cavity wall playing the role of the inversion sphere:
\begin{align}
\bm{p}_{2} = \bm{O} + \left(\frac{R}{|\bm{p}_{1}  - \bm{O} |}\right)^2(\bm{p}_{1}  - \bm{O})
\,\,.
\label{sphere_inversion_2}
\end{align}  

%
Let $r_{1}$ and $r_{2}$ be the distances between a point on the cavity surface and the points $\bm{p}_{1}$ and $\bm{p}_{2}$, respectively. 
For any point on the cavity surface, it can be proven that
\begin{align}
\frac{r_{1}^2}{|\bm{p}_{1}-\bm{O}|} = \frac{r_{2}^2}{|\bm{p}_{2}-\bm{O}|}
\,\,.
\label{governing_property_of_inversion}
\end{align}
From this property, it follows immediately that if the value of the image charge is assigned as 
\begin{align}
q_{2} = -\sqrt{\frac{|\bm{p}_{2}-\bm{O}|}{|\bm{p}_{1}-\bm{O}|}} q_{1}\,\,,
\label{charge_transformation__3D}
\end{align}
then the electrostatic potential created by the physical charge and its image 
on any point on the wall will vanish: $q_{1}/r_{1}+q_{2}/r_{2} = 0$ The 
resulting field will 
be the correct solution of the Poisson equation with (zero) Dirichlet boundary 
conditions, thus solving the problem of the field induced by a charge. 

\section{Piecewise-spherical conducting cavities solvable with the method of images}
As we have been saying, our solvable problems will involve regions (cavities) bounded by segments of spheres and planes, where the latter coincide with the 4D$\to$3D stereographic projections of great hypercircles associated with a finite reflection group in 4D. Let us, however, first explicitly write down the conditions that must be satisfied in order for any problem involving regions bounded by grounded conducting segments of spheres and planes to be solvable by the method of (finitely many) images.

Imagine an empty cavity surrounded by a grounded conductor. Assume that its walls are formed by segments of spherical surfaces. The field induced by a point charge placed inside the cavity can be constructed using the method of images if the the system satisfies the following conditions, which are spherical-geometry analogs of the requirements described in Sec.~\ref{solvable_planar_problem}. Note that if a surface is a plane rather than a sphere, in what follows, a sphere inversion should be replaced by a simple reflection.

\vspace{\baselineskip}
\begin{minipage}{0.9\columnwidth}
\textbf{Solvability conditions:}
\begin{itemize}
\item[I.  ] Consider the set of image charge locations produced via chains of inversions from Eq.~(\ref{sphere_inversion_2}) with respect all the spheres, one sphere at each link of the chain. For any starting location, there should be a finite number of image locations such that once they are all in the set, further inversions of its elements will only produce locations that are already in the set. 
\item[II. ] One should be able to unambiguously assign the values of the image charges via a sequential application of the rule in Eq.~(\ref{charge_transformation__3D}). 
The assigned charge value should not depend on the particular sequence of inversions that produced its location.
\item[III.] The sphere inversions should not produce any image charges within the cavity of interest.
\end{itemize}
\end{minipage}
\vspace{\baselineskip}

Let us show that these conditions are indeed sufficient. Consider one of the spherical surfaces defining the cavity. According to condition I., any charge---regardless of whether it is the physical charge or an image charge---will have a counterpart 
linked to it, in both directions, by an inversion with respect to the surface of interest. According to condition II., the two will create a zero potential on the cavity surface. The rest of the charges will form similar pairs, with similar results. Finally, condition III. guarantees that no `ghost' charges are required by the 
solution obtained, i.e.\ that inside the cavity there is only the charge that is present in the statement of the problem.

\section{Generation of piecewise-spherical conducting cavities using a 4D reflection group and a 4D$\to$3D stereographic projection}
\subsection{The setup}
\begin{figure}[!ht]
\centering
\includegraphics[width=0.9\columnwidth]{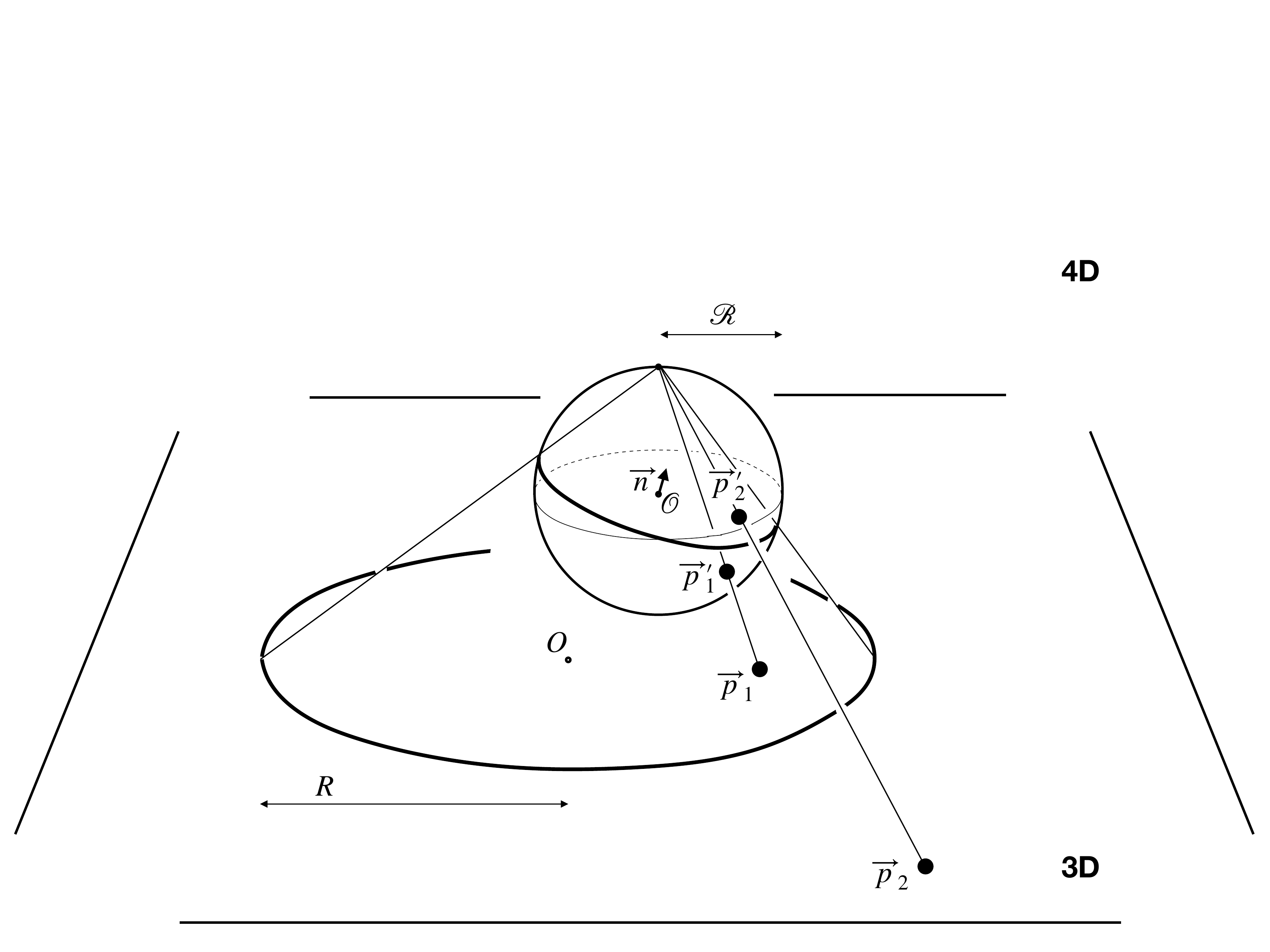}
\caption{
An artistic rendering of the relationship between a 4D reflection on a hypersphere and a 3D sphere inversion, through a 4D stereographic
projection.  The points $\bm{p}'_{1}$ and $\bm{p}'_{2}$ are on a hypersphere of radius $\mathcal{R}$ whose center is at $\bm{\mathcal{O}}$. These two points are related by a 4D reflection with respect to a hyperplane through $\bm{\mathcal{O}}$ whose unit normal vector is $\bm{n}$. The 3D points $\bm{p}_{1}$ and $\bm{p}_{2}$  are the images of $\bm{p}'_{1}$ and $\bm{p}'_{2}$ under the stereographic projection from the hypersphere to a `horizontal' hyperplane. The latter hyperplane is identified with the physical 3D space. It can be shown that $\bm{p}_{1}$ and $\bm{p}_{2}$ are 
then related via a 3D sphere inversion with respect to a sphere whose center is $\bm{O}$ and whose radius is $R$. 
This sphere is the stereographic image of the great hypercircle formed at the intersection of the hyperplane characterized by $\bm{n}$ and the hypersphere.  
            }
\label{f:4Dto3D}        
\end{figure}

As we have been suggesting, a fruitful way to construct a set of spherical surfaces that satisfy the solvability conditions outlined in the previous section is to use the construction described in Sec.~\ref{4D_outline}. That construction involves a 4D$\to$3D stereographic projection of great hypercircles on the surface of a 4D hypersphere, so let us describe this transformation quantitatively; see Fig.~\ref{f:4Dto3D}. 
A 4D$\to$3D stereographic projection takes a point $ \bm{\mathcal{p}}'$ on a hypersphere centered at $\bm{\mathcal{O}}$ and of radius $\mathcal{R}$, 
\begin{equation}
 \bm{\mathcal{p}}'\equiv\left(\begin{array}{c}p'_{x}\\p'_{y}\\p'_{z}\\p'_{w}\end{array}\right)=\left(\begin{array}{c}
                                                                         \sin(\Theta_{\bm{p}'})\sin(\theta_{\bm{p}'})\cos(\phi_{\bm{p}'}) \mathcal{R}
                                                                         \\
                                                                         \sin(\Theta_{\bm{p}'})\sin(\theta_{\bm{p}'})\sin(\phi_{\bm{p}'}) \mathcal{R}
                                                                         \\
                                                                         \sin(\Theta_{\bm{p}'})\cos(\theta_{\bm{p}'}) \mathcal{R}
                                                                         \\
                                                                         \cos(\Theta_{\bm{p}'}) \mathcal{R}
 \end{array}\right)
\,\,.
\label{stereographic_projection_origin}
\end{equation}
and converts it to a point $ \bm{\mathcal{p}}$ on the tangential `horizontal' 4D hyperplane---identified with the physical 3D space---that is tangent to the hypersphere at the `South Pole'
$(x,\,y,\,z,\,w)=(0,\,0,\,0,\,-\mathcal{R})$. In Cartesian coordinates, $(p'_{x},\,p'_{y},\,p'_{z},\,p'_{w})$ on the surface of the hypersphere gets mapped to $(p_{x},\,p_{y},\,p_{z},\,p_{w})=\frac{2\mathcal{R}}{\mathcal{R}-p'_{w}}\,(p'_{x},\,p'_{y},\,p'_{z},\,0)+(0,\,0,\,0,\,-\mathcal{R})$. In spherical coordinates,
\begin{equation}
 \bm{\mathcal{p}}\equiv\left(\begin{array}{c}p_{x}\\p_{y}\\p_{z}\\p_{w}\end{array}\right)=
                                                                 \left(\begin{array}{c} 
                                                                         2 \cot(\frac{\Theta_{\bm{p}'}}{2}) \sin(\theta_{\bm{p}'})  \cos(\phi_{\bm{p}'})        \mathcal{R}
                                                                         \\
                                                                         2 \cot(\frac{\Theta_{\bm{p}'}}{2}) \sin(\theta_{\bm{p}'})  \sin(\phi_{\bm{p}'})        \mathcal{R}
                                                                         \\
                                                                         2 \cot(\frac{\Theta_{\bm{p}'}}{2}) \cos(\theta_{\bm{p}'})                                      \mathcal{R}
                                                                         \\
                                                                         -\mathcal{R}
                                                                   \end{array}\right)
\,\,.
\label{stereographic_projection}
\end{equation}
The final step is an orthogonal projection $\proj$ to the 3D space, which is accomplished by simply deleting the final ($w$) coordinate: $\proj\, (x,\,y,\,z,\,w)=(x,\,y,\,z)$. Since it is unlikely to cause confusion, we will not always carefully distinguish the points before and after the orthogonal projection.

The stereographic projection is invertible: the 3D point $(p_{x},\,p_{y},\,p_{z})$ has the inverse image 
%
%
\[
\begin{pmatrix}p'_{x}\\p'_{y} \\p'_{z}\\p'_{w}
\end{pmatrix}
=
\frac{4\mathcal{R}^2}{4\mathcal{R}^2+p_{x}^{2}+p_{y}^{2}+p_{z}^{2}}
\begin{pmatrix}
 p_{x}\\p_{y}\\p_{z}\\-2\mathcal{R}
\end{pmatrix}
+
\begin{pmatrix}
 0\\0\\0\\ \mathcal{R}
\end{pmatrix}\,.
\]
%

A great hypercircle on our hypersphere is the set of 4D points that satisfy
%
\begin{eqnarray}
&
( \bm{\mathcal{p}}')^2 = \mathcal{R}^2
\label{great_hyper-circle__sphere}
\\
&
 \bm{\mathcal{p}}'\cdot  \bm{\mathcal{n}} = 0
\label{great_hyper-circle__plane}
\,\,,
\end{eqnarray}
%
where
\begin{align}
 \bm{\mathcal{n}}\equiv\left(\begin{array}{c}n_{x}\\n_{y}\\n_{z}\\n_{w}\end{array}\right)=\left(\begin{array}{c}  
                                                                         \sin(\Theta_{\bm{n}})\sin(\theta_{\bm{n}})\cos(\phi_{\bm{n}})
                                                                         \\
                                                                         \sin(\Theta_{\bm{n}})\sin(\theta_{\bm{n}})\sin(\phi_{\bm{n}})
                                                                         \\
                                                                         \sin(\Theta_{\bm{n}})\cos(\theta_{\bm{n}})
                                                                         \\
                                                                         \cos(\Theta_{\bm{n}})
                                                                  \end{array}\right)
\,\,.                                                                   
\label{4D_great_circle_normal}
\end{align}
We may interpret $ \bm{\mathcal{n}}$ as the unit vector normal to a hyperplane that passes through $\bm{\mathcal{O}}$, the center of the hypersphere. The intersection of this hyperplane and the hypersphere is the great hypercircle of interest.

Under the stereographic projection described by Eqs.~(\ref{stereographic_projection_origin}) and (\ref{stereographic_projection}), 
the great hypercircle defined by Eqs.~(\ref{great_hyper-circle__sphere}-\ref{4D_great_circle_normal}) transforms to a 3D sphere of radius 
\begin{align}
R = 2|\sec(\Theta_{\bm{n}})| \mathcal{R}
\,\,
\label{3D_inversion_sphere_radius}
\end{align}
centered at 
\begin{align}
\bm{\mathcal{O}}\equiv\left(\begin{array}{c}O_{x}\\O_{y}\\O_{z}\\O_{w}\end{array}\right)=
                                                                \left(\begin{array}{c}  
                                                                         -2 \tan(\Theta_{\bm{n}}) \sin(\theta_{\bm{n}})  \cos(\phi_{\bm{n}})        \mathcal{R}
                                                                         \\
                                                                         -2 \tan(\Theta_{\bm{n}}) \sin(\theta_{\bm{n}})  \sin(\phi_{\bm{n}})         \mathcal{R}
                                                                         \\
                                                                         -2 \tan(\Theta_{\bm{n}}) \cos(\theta_{\bm{n}})                                       \mathcal{R}
                                                                         \\
                                                                         -\mathcal{R}
                                                                   \end{array}\right)
\,\,.
\label{3D_inversion_sphere_center_in_4D}
\end{align}
Notice that all the spheres produced this way satisfy the relation
\begin{align}
R^2 = \bm{O}^2 + 4 \mathcal{R}^2
\,\,,
\label{great_circle_constraint}
\end{align}
with $\bm{O}$ being the 3D version of the corresponding 4D vector $\bm{\mathcal{O}}$, i.e.\ its orthogonal projection $\proj$ to the $xyz$ space:
\begin{align}
\bm{O} \equiv \proj \bm{\mathcal{O}} = (O_{x},\,O_{y},\,O_{z})
\label{3D_inversion_sphere_center_in_3D}
\,\,.
\end{align}  

Let $\bm{p}'_{1,\,\text{4D}}$ and $\bm{p}'_{2,\,\text{4D}}$ be two points on the hypersphere. Suppose they are related by a reflection via a 4D mirror defined by the unit normal vector $ \bm{\mathcal{n}}$, in other words that
\begin{align*}
\bm{p}'_{2,\,\text{4D}} = \bm{p}'_{1,\,\text{4D}} - 2( \bm{\mathcal{n}}\cdot\bm{p}'_{1,\,\text{4D}} ) \bm{\mathcal{n}}
\,\,.
\end{align*}
As we already mentioned, it can be explicitly shown that the stereographic images of $\bm{p}'_{1,\,\text{4D}}$ and $\bm{p}'_{2,\,\text{4D}}$ are related by a sphere inversion. To apply Eq.~(\ref{sphere_inversion_2}), we take $\bm{p}_{1}=\proj \bm{p}_{1,\,\text{4D}}=(p_{1,\,x},\,p_{1,\,y},\,p_{1,\,z})$ and similarly $\bm{p}_{2}=\proj \bm{p}_{2,\,\text{4D}}$. The inversion sphere parameters $R$ and $\bm{O}$ are as in Eqs.~(\ref{3D_inversion_sphere_radius}) and (\ref{3D_inversion_sphere_center_in_3D}).

Let  us finally consider the 3D spheres forming the cavity of interest. Let us assume that these 3D spheres stereographically originate from great hypercircles that are, in turn, produced by the 
generating mirrors of a 4D reflection group, as explained in Sec.~\ref{4D_outline}. 

\subsection{Solvability condition I: finite number of image charges}
Let us place a charge $q$ at the location $\bm{p}$ inside the cavity described above. Let $ \bm{\mathcal{p}}'$ be the stereographic inverse image of $\bm{p}$  (so that $\bm{p}$ is the stereographic image of $ \bm{\mathcal{p}}'$). Start iteratively reflecting $ \bm{\mathcal{p}}'$ through the generating mirrors of the 4D finite reflection group. For each pair of 4D positions related by a reflection, their stereographic projections are related by a sphere inversion, and vice versa. Since  we know there are only finitely many images produced by the 4D reflections, it follows  that the iterative sphere inversions can produce only a finite number of image charges (the same number as the 4D reflections produce).

\subsection{Solvability condition II: consistency of charge assignment}
This is a sphere-inversion analog of the discussion in Sec.~\ref{4D_outline} about how to assign charge values to the image charges in the case of mirror reflections. But this is one place where the sphere inversion is noticeably more complicated than its 4D mirror-reflection counterpart, because while the reflections only affect the signs of the image charges, sphere inversions also affect their magnitudes.

If an inversion sphere is a stereographic image of a great hypercircle on a hypersphere of radius $\mathcal{R}$, then any two points $\bm{p}_{1}$ and $\bm{p}_{2}$ related by the 
inversion obey
\begin{align}
\frac{|\bm{p}_{1}-\bm{O}|}{|\bm{p}_{2}-\bm{O}|} = \frac{F(\bm{p}_{1})}{F(\bm{p}_{2})}
\label{charge_connection_1}
\,\,,
\end{align}  
with
\begin{align}
F(\bm{p}) = \bm{p}^2 + 4 \mathcal{R}^2
\label{charge_connection_2}
\,\,.
\end{align}  
Notice that $F(\bm{p}_{1})$ depends neither on the other point $\bm{p}_{2}$, nor on the center $\bm{O}$ of the inversion sphere, nor on the radius $R$ of the inversion sphere. The same holds for 
$F(\bm{p}_{2})$, upon a $\bm{p}_{1} \leftrightarrow \bm{p}_{2}$ substitution. The charge assignment rule in Eq.~(\ref{charge_transformation__3D}) becomes
\begin{align}
q_{2} = -\sqrt{\frac{F({\bm{p}_{2}})}{F(\bm{p}_{1})}} q_{1}
\,\,.
\label{charge_transformation__3D__special}
\end{align}
It is easy to show that this relationship will become
\begin{align}
q_{1+m} = (-1)^{m}\sqrt{\frac{F({\bm{p}_{2}})}{F(\bm{p}_{1})}} q_{1}
\,\,
\label{charge_transformation__3D__special_2}
\end{align}
if $q_{1+m}$ and $q_{1}$ are linked by a chain of $m$ inversions (instead of a single inversion). In particular, any image charge $q_{\text{image}}$ will be 
related to the original physical charge $q_{\text{physical}}$ as 
\begin{align}
q_{\text{image}} = (-1)^{m_{\text{image}}}\sqrt{\frac{F({\bm{p}_{\text{image}}})}{F(\bm{p}_{\text{physical}})}} q_{\text{physical}}
\,\,,
\label{charge_transformation__3D__special_3}
\end{align}
where $(-1)^{m_{\text{image}}}$ is the parity of the number of inversions linking the physical charge and the image charge in question. As we already mentioned in Sec.~\ref{4D_outline}, it is known that for reflection groups, the parity of the number of reflections leading to a particular member of the group is the same for any chain of reflections. 
Hence, the same will be true for sphere inversions that are stereographically linked to the members of a 4D reflection group, and the charge value 
assigned using Eq.~(\ref{charge_transformation__3D__special_3}) is indeed independent of the sequence of inversions used to get from the original charge to the image charge.

\subsection{Solvability condition III: no image charges outside of the conductor}
Consider again $ \bm{\mathcal{p}}'$, the stereographic inverse image of $\bm{p}$, where the latter is the location of the physical charge. We have already seen above that there will be one image charge per non-principal chamber, and so in particular all the 4D mirror-reflection images of $ \bm{\mathcal{p}}'$ will lie outside of the principal chamber. The intersection of the principal chamber with the hypersphere will, under the 
stereographic projection, become the 3D cavity of interest. Likewise, the 4D reflection images will become the locations of the 3D image charges, and none of them will be located inside the cavity of interest.

\section{An example: a solvable cavity associated with the reflection group $\gr{D}{4}$}

According to Table~\ref{reflection_groups_1}, the 4D reflection group $\gr{D}{4}$ has the Coxeter diagram 
\dynkin[Coxeter]{D}{4}\,. 
Thus it has four simple roots (and so four generating mirrors), three of which are mutually orthogonal, while the fourth one makes an angle of $\pi/3$ with each of the first three.  The total number of group elements is is $192$, of which $12$ are pure reflections. Once the mutual orientation of the generating mirrors is fixed (basically by the Coxeter diagram), we have to choose the orientation 
of the mirrors as a group. The point where all the generating mirrors meet should be fixed at the origin, so the allowed transformations are 4D rotations. In four dimensions, rotations are parametrized by $6$ real parameters. Indeed, the full reflection group can be parametrized as a product six rotation matrices \cite{Hanson_1994}, each describing a rotation in one of six pairs of Cartesian planes: $xy$, $xz$, $yz$, $xw$, $yw$, and $zw$. We will refer to these as \emph{elementary rotations}. For example, the matrix describing an elementary rotation of a 4D vector $(x,\,y,\,z,\,w)$ that takes place in the $yw$-plane is 
\[
R_{yw}(\theta_{yw})=\begin{pmatrix}1 & 0 & 0 & 0\\
0 &  \cos \theta_{yw} & 0 & \sin \theta_{yw} \\
0 & 0 & 1 & 0\\
0 &  -\sin \theta_{yw} &0 & \cos \theta_{yw} \\                                                                                                                                                                                                                                                                                                                                                               \end{pmatrix}\,.                                                                                                                                                                                                                   \]
The other elementary rotations work similarly, where each one gets its own angle (e.g. $\theta_{xy}$, $\theta_{xz}$, etc.). All six angles can be set independently.

Note that the concept of a rotation axis is not useful in dimensions higher than three, where the dimensionality of what is left fixed by a rotation in general depends on the rotation---unlike in 3D, where every rotation leaves fixed a line trough the origin. In 4D there are rotations that, like our elementary ones, leave a whole 2D plane fixed. But the most general 4D rotation is a double rotation, a simultaneous rotation in two perpendicular planes, which leaves only the origin fixed. 

Recall that our stereographic projection maps into the $xyz$ space. Let us parametrize an arbitrary 4D rotation by first performing elementary rotations in planes involving the fourth dimension $w$, and then performing the rotations that take place wholly within the $xyz$ space. For example, we can use the matrix 
$R_{yz}R_{xz}R_{xy}R_{zw}R_{yw}R_{xw}$, where $R_{xw}$ is a function of $\theta_{xw}$, $R_{yw}$ of $\theta_{yw}$, etc.
In this case, the final three rotations ($R_{yz}R_{xz}R_{xy}$) will have the effect of simply 3D-rotate the final system of spheres and planes. The 3D$\to$2D analog is a rotation of the sphere in Fig.~\ref{tetrahedron}(b) about the $z$-axis. In Fig.~\ref{tetrahedron}(c) and (d), this would have the effect of rigidly rotating the whole plane about the South Pole. 

The first three rotations, however, can be used to control which great hypercircles stereographically map into planes and which into spheres, and what are the radii of the spheres and the locations of spheres and planes. The 3D$\to$2D analog is rotating the sphere in Fig.~\ref{tetrahedron}(b) about the $x$ or $y$ axis. This will generally deform the patterns we get in the plane, changing the radii of the spheres and possibly even transforming them into planes.

As an example, we choose the following set of the generating mirror normals:
\begin{align}
&
\begin{array}{c}
(1,\,0,\,0,\,0)
\\
(0,\,1,\,0,\,0)
\\
(0,\,0,\,1,\,0)
\end{array}
\label{D_4_normals_123}
\\
&
\left(\frac{1}{2},\,\frac{1}{2},\,\frac{1}{2},\,\frac{1}{2}\right)
\label{D_4_normal_4}
\,\,.
\end{align}
The corresponding angles are then given by Eq.~(\ref{4D_great_circle_normal}), which gives
\begin{align}
&
\begin{array}{l}
\Theta = \frac{\pi}{2};\, \theta = \frac{\pi}{2};\, \phi = 0
\\
\Theta = \frac{\pi}{2};\, \theta = \frac{\pi}{2};\, \phi = \frac{\pi}{2}
\\
\Theta = \frac{\pi}{2};\, \theta = 0;\, \phi = \text{any}
\end{array}
\label{D_4_angles_123}
\\
&
\begin{array}{l}
\Theta = \frac{\pi}{3};\, \theta = \arg\left(\frac{1 + i \sqrt{2}}{\sqrt{3}}\right);\, \phi = \frac{\pi}{4}
\end{array}
\label{D_4_angles_4}
\,\,.
\end{align}

Under a stereographic projection  in Eqs.~(\ref{stereographic_projection_origin}) and (\ref{stereographic_projection}), Eq.~(\ref{3D_inversion_sphere_radius}) says that the first three great hypercircles 
transform to 3D spheres of infinite radius, i.e.\ to 3D planes. These planes will cross the origin. The fourth 
hypercircle becomes a sphere. The resulting cavity is given by the following set of inequalities:
\begin{align}
x>0,\,y>0,\,z>0
\,\,,
\label{D_4_planes}
\end{align}
and
\begin{align}
\left(x+\frac{R}{2}\right)^{2}+\left(y+\frac{R}{2}\right)^{2}+\left(z+\frac{R}{2}\right)^{2}<R^2
\,\,,
\label{D_4_sphere}
\end{align}
with
\begin{align}
R = 4\mathcal{R}
\,\,,
\label{D_4_sphere_radius}
\end{align}
where $\mathcal{R}$ is the radius of the hypersphere the stereographic projection originates from;
see Eqs.~(\ref{3D_inversion_sphere_center_in_3D}), (\ref{3D_inversion_sphere_center_in_4D}), and (\ref{3D_inversion_sphere_radius}) 
for the formulae for the center and the radius of the ball defined by Eq.~(\ref{D_4_sphere}).
Notice that since the stereographic projection is a conformal transformation, the angles between the resulting 3D surfaces are the same as the angles between the 4D 
hyperplanes from Eqs.~(\ref{D_4_normals_123}) and (\ref{D_4_normal_4}) that generate them. In particular, the spherical segment of the cavity surface crosses each of the three planar
segments at an angle of \ang{60}.

We can use the method of images to find the electrostatic potential created by a point charge placed anywhere inside the cavity bounded by the surfaces defined in Eqs.~(\ref{D_4_planes}) and (\ref{D_4_sphere}), where the surfaces are grounded conducting walls. 
The image charge locations will be given by sequential applications of the reflections about 
the planar cavity boundaries and sphere inversions in Eq.~(\ref{sphere_inversion_2}).
The number of image charges, all situated outside the cavity, is $191$. The charge values can be unambiguously assigned using the rule in Eqs.~(\ref{charge_transformation__3D__special_3}) and (\ref{charge_connection_2}), indeed the original rule in Eq.~(\ref{charge_transformation__3D}). For the reflections, when an inversion 
sphere degenerates into a plane,
the rules in Eqs.~(\ref{charge_transformation__3D__special_3}), (\ref{charge_connection_2}) and (\ref{charge_transformation__3D}) become 
$
q_{\text{image}} = (-1)^{m} q_{\text{physical}}
$
and
$
q_{2} = - q_{1}
$, 
respectively,
with $m$ still being the total number of the inversions and reflections linking $q_{\text{image}}$ and $q_{\text{physical}}$.
\begin{figure}[!ht]
\centering
\includegraphics[width=\columnwidth]{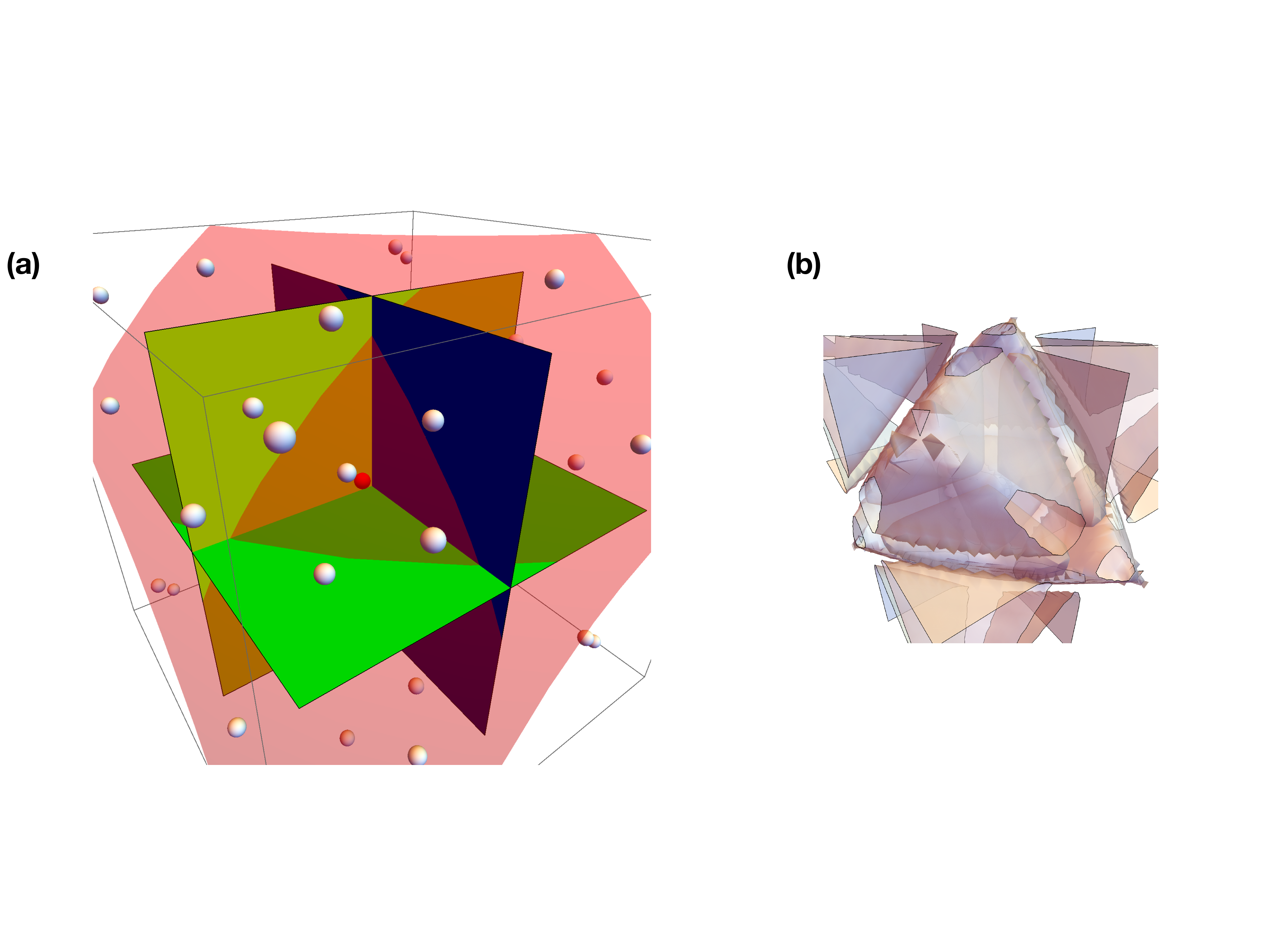}
\caption{A solvable electrostatic problem associated with the four-dimensional reflection group $\gr{D}{4}$ (a subgroup of the full symmetry group of the tesseract).
(a) A cubic segment of space, centered at the origin, of a linear size $0.6 R$.  (b) The surfaces of zero potential for the electrostatic potential induced by the sample charge of the previous subfigure.
             }
\label{f:D_4__COMPOUND}        
\end{figure}

In Fig.~\ref{f:D_4__COMPOUND}(a), we show a cubic segment of space, centered at the origin, of a linear size $0.6 R$. The tetrahedron-like shape in the middle of the subfigure represents the shape of an electrostatic cavity---with grounded conducting walls---solvable using the method 
of images. A sample charge (red) is placed at the grand diagonal, at a distance $0.05 R$ away from the spherical segment of the surface of the cavity. (Solvability persists for any position of the charge inside the cavity.) There are 191 image charges, but only a subset of them are visible (white) in the subfigure. 
The total number of charges, corresponding to the one physical charge supplemented by all the image charges, is $192$, which is also the total number of group elements---both reflections and rotations---in the reflection group $\gr{D}{4}$.  Together with the sample (physical) charge, 
the image charges generate a potential inside the cavity that vanishes at the cavity walls.

Figure~\ref{f:D_4__COMPOUND}(b) presents the surfaces of zero potential for the electrostatic potential induced by the sample charge of the previous subfigure. Shown there is a cubic segment of space, centered at the origin, of a linear 
size $0.36 R$. A part of this zero-potential surface coincides with the surface of the cavity. The view of that part is partially obscured by the eight additional spheres (additional to the four surfaces forming the cavity) on which the potential vanishes as well. There must be a total of 12 surfaces, each either a sphere or a plane, because \gr{D}{4} has a total of 12 mirrors, and the symmetry of the problem guarantees they all must have the same potential. The potential also vanishes on the continuations of the surface fragments forming the cavity of interest. 
\section{Summary and outlook}

In all branches of physical science, problems whose solution is expressible in closed form are invaluable for the development of physical insight, while also serving as benchmarks on which numerical or approximate (but more generally applicable) methods of solution can be tested  \cite{Palaniappan_2011_107,sutherland2004_book,albeverio_solvable_2005,ablowitz1981}. In some dynamical cases, exact solvability is itself the mechanism underlying important physical effects (such as lack of thermalization \cite{Kinoshita2006_440}) and distinct physical phenomena (such as solitons \cite{ablowitz1981}). This is due to the fact that exactly solvable (also known as \emph{integrable}) time-dependent systems have a great (often infinite) number of dynamical conservation laws.

We have shown that in electrostatics, the class of problems solvable by the method of images is considerably larger than what was previously known. In particular, we showed that the solution of a classic electrostatic problem---finding the field induced by a point charge placed inside a spherical cavity with grounded conducting walls---can be generalized to any cavity whose walls are represented by 4D stereographic projections of great hypercircles formed by the intersections of a 4D hypersphere and the mirrors of any known finite 4D reflection group of rank 4 or lower. We used the group $\gr{D}{4}$ as a worked example. 

In the process, we pointed out that every known problem that is solvable by adding finitely image charges is based on some finite reflection group. Starting with such a group, one may use its mirrors as they are or apply further transformations to them: either a 4D$\to$3D stereographic projection, which we introduced in this work, or the well-known Kelvin transform. In this way we obtain all presently known solvable problems. 

The scope of problems solvable by our construction is as follows:
\begin{itemize}
\item[$\bullet$]
Any finite reflection group  of rank 4 or less can be used, including groups that are reducible and those that are lower-dimensional but embedded in 4D; see Tables~\ref{reflection_groups_1} and \ref{reflection_groups_2}. 

\item[$\bullet$]
For each of the reflection groups listed in Tables~\ref{reflection_groups_1} and \ref{reflection_groups_2}, there will be a three-parametric family of orientations of the mirrors in the 4D space, leading to a three-parametric family of 3D sphere radii and positions. The remaining three parameters of the six-parametric family of the 4D rotations will control the trivial 3D rotation of the resulting 3D cavity.
\end{itemize}

A 2D generalization of our scheme (a 3D stereographic projection from a sphere to a plane) is worth considering. There, one obtains a family of solvable problems involving cylindrical conducting cavities and charged wires. The problem of assigning the values of charges is expected to disappear (as one trivially gets a set of sign-alternating charges). However, this property alone only guarantees that the electrostatic potential will be a non-zero constant on each of the cylindrical segments; it is still in principle possible that the potential will not be given by the same constant on each of the cylinders. However, we expect that a relationship analogous to Eqs.~(\ref{charge_connection_1}) and (\ref{charge_connection_2}) can be proven in the 2D case as well. 
If so, then it will be possible to prove that the potential in fact has the same value at every point on the cavity surface. 

The property in Eqs.~(\ref{charge_connection_1}) and (\ref{charge_connection_2})---that so far seems completely accidental---may prove to be a consequence of a deeper connection. 
One may conjecture that a map exists between the solutions of the Poisson equation on the hypersphere and the solutions of the Poisson equation on a plane stereographically connected 
to that hypersphere. In this case, the solutions for the class of problems we considered will become images of the hyperspherical solutions which were obtained via \emph{pure reflections}, obviating the need to go through sphere inversion procedure in 3D. 
The charges on the hypersphere will have the same magnitude and an alternating sign. The relationship between a hyperspherical charge $q'$ and the 3D charge 
$q$ will be then given by $q = \text{const} \times \sqrt{F(\bm{p})} \, q'$, where $\bm{p}$ is the location of the charge $q$, and the function $F$ is given by Eq.~(\ref{charge_connection_2}). The proof of this conjecture may involve (a) the fact that the 3D sphere-, and, potentially, 4D hypersphere inversions convert solutions of the Poisson equation to solutions of the Poisson equation and (b) the fact that a 4D stereographic projection can be reinterpreted as a 4D hypersphere inversion.

Finally, it is hard to believe that our construction is not in some way connected to the Kelvin transform after all. A progress regarding the program from the previous paragraph might help elucidate this connection, and perhaps the other way around as well.

\providecommand{\noopsort}[1]{}\providecommand{\singleletter}[1]{#1}%


\begin{thebibliography}{10}
\providecommand{\url}[1]{{#1}}
\providecommand{\urlprefix}{URL }
\expandafter\ifx\csname urlstyle\endcsname\relax
  \providecommand{\doi}[1]{DOI \discretionary{}{}{}#1}\else
  \providecommand{\doi}{DOI \discretionary{}{}{}\begingroup
  \urlstyle{rm}\Url}\fi

\bibitem{landauEandMinmedia}
L.D. Landau, L.P. Pitaevskii, \emph{Electrodynamics of Continuous Media: v.~8},
  2nd edn. (Pergamon, New York, 1984).
\newblock \doi{10.1016/B978-0-08-030275-1.50025-4}

\bibitem{jackson_E_and_M}
J.D. Jackson, \emph{Classical Electrodynamics}, 3rd edn. (Wiley, Hoboken, NJ,
  2009)

\bibitem{Palaniappan_2011_107}
D.~Palaniappan, Electr. Eng. \textbf{94}, 107 (2011).
\newblock \doi{10.1007/s00202-011-0221-7}

\bibitem{Garg_2012}
A.~Garg, \emph{Classical electromagnetism in a nutshell} (Princeton University
  Press, Princeton, NJ, 2012)

\bibitem{Smythe_1989}
W.R. Smythe, \emph{Static and Dynamic Electricity}, 3rd edn. (Hemisphere Pub.
  Corp., New York, 1989)

\bibitem{Jeans_2009}
J.H. Jeans, \emph{The Mathematical Theory of Electricity and Magnetism},
  reprinted 5th edn. (Cambridge University Press, Cambridge, UK, 2009).
\newblock \doi{10.1017/CBO9780511694356}

\bibitem{Maxwell_1873}
J.C. Maxwell, \emph{A Treatise on Electricity and Magnetism} (Macmillan,
  London, UK, 1873), vol.~1, chap.~XI

\bibitem{Humphreys_1994}
J.E. Humphreys, \emph{Reflection Groups and {C}oxeter Groups} (Cambridge
  University Press, Cambridge, U.K., 2012).
\newblock \doi{10.1017/CBO9780511623646}

\bibitem{Olshanii_2015_105005}
M.~Olshanii, S.G. Jackson, New J. Phys. \textbf{17}, 105005 (2015).
\newblock \doi{10.1088/1367-2630/17/10/105005}

\bibitem{gaudin1983_book_english}
M.~Gaudin, J.S. Caux, \emph{The {B}ethe wavefunction} (Cambridge University
  Press, Cambridge, UK, 2014)

\bibitem{sutherland2004_book}
B.~Sutherland, \emph{Beautiful Models: 70 Years of Exactly Solved Quantum
  Many-Body Problems} (World Scientific, Singapore, 2004).
\newblock \doi{10.1142/5552}

\bibitem{Rosenfeld_1977}
B.A. Rosenfeld, N.D. Sergeeva, \emph{Stereographic Projection} (Mir Publishers,
  Moscow, Russia, 1977).
\newblock Translated from the {R}ussian by {V}italy {K}isin

\bibitem{Kellogg_1967}
O.D. Kellogg, \emph{Foundations of Potential Theory} (Springer, Berlin, 1967).
\newblock \doi{10.1007/978-3-642-86748-4}

\bibitem{Hanson_1994}
A.J. Hanson, in \emph{Graphics Gems IV}, ed. by P.S. Heckbert (Academic Press,
  San Diego, CA, USA, 1994), chap. II.6, pp. 149--170.
\newblock \doi{10.1016/B978-0-12-336156-1.50024-0}

\bibitem{albeverio_solvable_2005}
S.~Albeverio, F.~Gesztesy, R.~H{\o}egh-Krohn, H.~Holden, \emph{Solvable Models
  in Quantum Mechanics}, 2nd edn. (AMS Chelsea Pub., Providence, RI, 2005).
\newblock With an appendix by Pavel Exner

\bibitem{ablowitz1981}
M.J. Ablowitz, H.~Segur, \emph{Solitons and the Inverse Scattering Transform}
  (SIAM, Philadelphia, PA, 1981).
\newblock \doi{10.1137/1.9781611970883}

\bibitem{Kinoshita2006_440}
T.~Kinoshita, T.~Wenger, D.S. Weiss, Nature \textbf{440}, 900 (2006).
\newblock \doi{10.1038/nature04693}

\end{thebibliography}
\end{document}